%% file: Albaneseetal24_final.tex
\documentclass[english, 12pt, a4paper]{article}
\usepackage{chngcntr}
\usepackage[symbol]{footmisc}
\usepackage{floatpag}
\floatpagestyle{empty}
\usepackage{sectsty}
\usepackage{float}
\providecommand{\noopsort}[1]{}
\usepackage{url}
\usepackage{hyperref}
\usepackage{pdflscape}
\usepackage{mathptmx}
\usepackage{changepage}
\usepackage[T1]{fontenc}
\usepackage{lscape}
\usepackage[english]{babel}
\usepackage{booktabs}
\usepackage{fontenc}
\usepackage{tikz}
\usetikzlibrary{decorations.pathreplacing}
\usepackage{threeparttable}
\usepackage{longtable}
\usepackage{amsmath} %for boldsymbol
\usepackage{amsfonts}
\usepackage{amssymb}
\usepackage{mathrsfs}
\usepackage{bbm}
\usepackage{url}
\usepackage{graphicx}

\usepackage[official]{eurosym} %for euro symbol
\usepackage{rotating}
\usepackage{t1enc}
\usepackage{xcolor}
\usepackage{caption}
\usepackage{subcaption}

\usepackage[round]{natbib}
\usepackage{fullpage}
\usepackage{scrextend}
\deffootnote{0.5em}{0em}{\textsuperscript{\thefootnotemark}\,}
\usepackage{ragged2e}
\usepackage{hyperref}
\usepackage[nameinlink,capitalise]{cleveref}

\crefname{figure}{Figure}{Figures}%
\hypersetup{
	colorlinks=true,% hyperlinks will be black
	linkcolor = blue,
	linkbordercolor=,% hyperlink borders will be red
	citecolor=blue,
	pdfborderstyle=,
	urlcolor=blue
}

\usepackage{soul}
\usepackage{titlesec}
\usepackage{lscape} 
\usepackage{array} 

\definecolor{darkgreen}{rgb}{0.0, 0.42, 0.24}

\definecolor{copper}{rgb}{0.72, 0.45, 0.2}

\usepackage{setspace}
\usepackage[margin=1in]{geometry}

\begin{document}
	\begin{singlespace}
		
		\title{\textbf{{\Large{Long-Term Effects of Hiring Subsidies for Low-Educated Unemployed Youths}}}\thanks{\noindent We acknowledge the financial support for this research project from the CORE program of the Luxembourg National Research Fund (FNR) (project number 11700060). This paper uses confidential data from the Crossroads Bank for Social Security (CBSS) (contract no. ART5/18/033). The data can be obtained by filing a request directly with CBSS (\url{https://www.ksz-bcss.fgov.be/en}). The authors are willing to assist. We thank Sylvain Klein for the provision of the commuting-time statistics. We are grateful to the editor, David Seim, and three anonymous reviewers for their constructive comments. We also thank Sam Desiere, Felix Stips, Kostantinos Tatsiramos, Bruno Van der Linden, and the participants at the 33rd Annual Conference of the European Association of Labour Economists (EALE), the Counterfactual Methods for Policy Impact Evaluation (COMPIE) Conferences in 2021 and 2022, the seminar at the Competence Centre on Microeconomic Evaluation, Joint Research Center (JRC), and the seminar at LISER for their valuable suggestions. E-mail addresses: \href{andrea.albanese@liser.lu}{andrea.albanese@liser.lu} (Andrea Albanese, corresponding author), \href{bart.cockx@ugent.be}{bart.cockx@ugent.be} (Bart Cockx), \href{muriel.dejemeppe@uclouvain.be}{muriel.dejemeppe@uclouvain.be} (Muriel Dejemeppe).

		}}	
		\author{Andrea Albanese$^{\textrm{a,b,c,d,e}}$, Bart Cockx$^{\textrm{b,c,d,f,g}}$, Muriel Dejemeppe$^{\textrm{c,e}}$ \\
			\footnotesize{$^{\textrm{a}}$ {\em Luxembourg Institute of Socio-Economic Research (LISER), Luxembourg}} \vspace{-.1cm}\\
			\footnotesize{$^{\textrm{b}}$ {\em Department of Economics, Ghent University, Belgium}} \vspace{-.1cm}\\
			\footnotesize{$^{\textrm{c}}$ {\em IRES/LIDAM, UCLouvain, Belgium}} \vspace{-.1cm}\\
			\footnotesize{$^{\textrm{d}}$ {\em IZA, Bonn, Germany}} \vspace{-.1cm}\\
			\footnotesize{$^{\textrm{e}}$ {\em GLO, Essen, Germany }}\vspace{-.1cm}\\
			\footnotesize{$^{\textrm{f}}$ {\em CESIfo, Munich, Germany }}\vspace{-.1cm}\\
			\footnotesize{$^{\textrm{g}}$ {\em ROA, Maastricht University }} \vspace{-.1cm}}
		
		\date{March 3, 2024}
		\maketitle
		\thispagestyle{empty}

		\abstract{\noindent We use regression discontinuity design and difference-in-differences methods to estimate the impact of a one-time hiring subsidy for low-educated unemployed youths in Belgium during the recovery from the Great Recession. Within a year of unemployment, the subsidy increases job-finding in the private sector by 10 percentage points. Over six years, high school graduates secure 2.8 more quarters of private employment. However, they transition from public jobs and self-employment, resulting in no net increase in overall employment, albeit with better wages. High school dropouts experience no lasting benefits. Additionally, in tight labor markets near Luxembourg's employment hub, the subsidy results in a complete deadweight loss. 
		}

	\end{singlespace}
	\setcounter{page}{0} % reset page counter 
	
	\newpage

	\clearpage
	\renewcommand*{\thefootnote}{\arabic{footnote}}
	\section{Introduction}\label{sec:introduction}

	Economic recessions generally affect the labor market position of young people more strongly than that of adults. Following the 2008 Great Recession and the European sovereign debt crisis, the youth unemployment rate in the EU-27 increased from 16.0\% in 2007 to 24.4\% in 2013. Low-educated youths (less than high school education) were particularly affected, facing an unemployment rate increase from 19.5\% to 31.0\% over the same period \citep{Eurostat23a}. The empirical literature has demonstrated that recessions can have long-lasting consequences on the careers of youths \citep[for example,][]{cockx2016, Wachter20, Wachter21}. To counter these impediments to successful careers for youths, various policy interventions are often proposed, such as training, job-search assistance, monitoring, and direct job creation (see, for example, \citealp{Caliendo16}). Finding the most appropriate policy responses is high on the policy agenda \citep{OECD20}. 
	
	One frequently used policy is hiring subsidies for low-educated youths, which are often advocated to counteract negative demand shocks. The canonical economic literature shows that hiring subsidies can stimulate new employment if labor supply and demand are sufficiently elastic;  otherwise, they are absorbed by higher wages \citep{Katz96}. Other factors may also influence the effectiveness of hiring subsidies, such as the extent to which they are targeted at specific groups (e.g., workers near the minimum wage; \citealp{Fougere00}),  are one-shot and unanticipated \citep{Cahuc19}, are implemented in a tight labor market \citep{Kline13}, or are conditional on job creation \citep{Neumark17}. As hiring subsidies have a limited duration, new job opportunities may be short-lived and have no effect beyond the expiration of the subsidy. However, employment gains can persist if a worker's productivity has had the opportunity to be unveiled or has grown enough to justify the increase in wage costs after the subsidy ends. First, firms may use the subsidy as a screening device to reveal a worker’s productivity and retain those who are highly productive \citep{Gibbons96, Altonji01, Pallais14}. Second, accumulation of firm-specific human capital during the subsidized period may raise productivity and encourage retention. Employment gains may also occur in other firms due to an accumulation of transferable human capital \citep{Acemoglu98, Acemoglu99, Autor01, Adda23}. A final factor shaping the effectiveness of hiring subsidies is the spillover effect they may trigger on ineligible individuals and the substitution of unsubsidized jobs (see, for example, \citealp{Crepon13}). 
	
	While the empirical literature has extensively studied hiring subsidies targeted at the long-term unemployed \citep{Schunemann15, Sjogren15, Ciani19, pasquini19, Desiere22}, much less is known about the effect of subsidies targeted at low-educated unemployed youths, in particular in the long run after the expiration of subsidies, and taking into account spillovers \citep{Dahlberg05}. 
	
	In this paper, we fill these gaps by studying the short- and long-run effects on various labor market outcomes of a generous one-shot hiring subsidy implemented in Belgium during the recovery from the Great Recession. This subsidy, called the \emph{Win-Win Plan}, targeted young low-educated unemployed jobseekers. A firm hiring a high school dropout younger than 26 was eligible for a monthly subsidy of \texteuro1,100, which represented 48\% of wage costs, on average. For high school graduates the amount was marginally lower and amounted to \texteuro1,000/month, i.e., 41\% of their wage costs. The subsidy was granted for two years for hirings in 2010 and for one year for those in 2011. Since other, less generous, hiring subsidies  existed for unemployed persons older than 26, the reform introduced an incremental reduction of wage costs of 24 percentage points for the high school dropouts and 18 percentage points for the graduates. 
	
	To estimate the intention-to-treat effect, we apply a one-sided donut regression discontinuity analysis \citep{Barreca16, Gerard21} that exploits the discontinuity in subsidy generosity at age 26. The estimation is based on a large sample of unemployed individuals living in the southern part of Belgium drawn from social security register data. We assess the robustness of our findings using the doubly robust semi-parametric difference-in-differences method of \citet{SANTANNA20}. 
	
	In a nutshell, we find no evidence of incidence of the subsidy reinforcement on wages. The transition to private sector employment is raised by 10 percentage points within the first year of unemployment. For high school dropouts, the positive effect is short-lived and does not persist beyond the end of the subsidy period. In contrast, for high school graduates this positive effect on private employment does persist. Seven years after entry into unemployment, high school graduates accumulated an average of 2.8 more quarters in private employment, which is a relative increase of 28\% with respect to the counterfactual of no Win-Win plan. Yet, this positive effect for high school graduates is counterbalanced by a decline in lower quality public employment as well as self-employment. Additionally, we demonstrate that labor market tightness induced by geographical proximity to the economic hub of Luxembourg diminishes the impact of the hiring subsidy. Lastly, we calculate the marginal value of public funds (MVPF) following the method proposed by \citet{Hendren20}, indicating that the hiring subsidy is potentially self-financing for high school graduates. However, this estimate is very imprecise, so we do not recommend basing policy recommendations on this finding.
	
	Our contribution to the literature is threefold. First, we provide new evidence on the effectiveness of a pure hiring subsidy targeted at low-educated young jobseekers. The added value of our analysis is that we consider outcomes up to five years after the expiration of the subsidy, longer than most existing studies, and that we apply different strategies for identifying the causal effects. Previous empirical evidence relating to similar target groups is inconclusive. The early studies estimating the effects of entering a subsidized job relied on the conditional independence assumption. In Sweden, \citet{Larsson03} finds that a hiring subsidy targeted at young high school graduates did not enhance the employment of participants up to two years after participation. \citet{Costa13} even show a negative effect of this policy after correcting for failures of the conditional independence assumption. In contrast, \citet{Caliendo11} report large positive effects on the employment probability up to five years after being hired in temporary subsidized jobs in Germany for low-educated youths. Other studies estimate the impact of youth hiring subsidies mixed with other policy interventions such as job counselling \citep{Blundell04, Dorsett06} or training \citep{Bell11, Brodaty11}. While the effects tend to be positive, it is difficult to pin them down to one specific intervention.\footnote{Another part of the literature evaluates the effect of payroll tax cuts targeted at all young employees and not only new hires. Wage subsidies are usually meant to boost the labor market integration of the target group structurally, and not as a temporary measure following an economic shock. Furthermore, as they are not targeted only at new hires, they tend to induce a strong deadweight \citep{Neumark13} and may therefore be an expensive way of boosting employment. \citet{Saez19} and \citet{Saez21} demonstrate that in Sweden, a wage subsidy in the form of a tax cut for businesses substantially increased youth employment, with the effect persisting three years after the tax cut was no longer in place (see also \citealp{Skedinger14, Egebark18}).} In a study similar to our paper, \citet{Cahuc19} find a large intention-to-treat effect on employment of a hiring subsidy targeted at low-wage workers in small firms that was introduced in France during the Great Recession. \citet{batut21} shows that the effect persists up to two years after its end. The Win-Win subsidy that we analyze shares many features of the French hiring subsidy as it is also an unanticipated one-shot intervention, although it is targeted at low-educated unemployed youths rather than low-paid jobs in general.
	
	Second, we provide evidence that hiring subsidies can have a positive long-term effect only in jobs where skill level exceeds a minimum threshold, i.e., a high school degree. This finding aligns with literature concluding "work-first” policies are ineffective for the lowest-skilled workers (i.e., high school dropouts) because the skill level of their jobs is too limited to generate significant human capital accumulation \citep{Meghir96, Card05, Blundell06}.\footnote{While several studies have challenged this view (e.g., \citealp{Dyke06, Autor10, Pallais14, Riddell20}), \citet{Autor17} argue that by focusing on the average effects of job placement programs some of these studies may mask considerable effect heterogeneity and high program failure rates, particularly among the most disadvantaged participants. Specifically, the authors do not find any significant effects of direct-hire and temporary help job placements in the US on employment or earnings for participants in the lower tail of the earnings distribution, while among higher potential earners only direct hires foster positive effects. Temporary-help placements even lead to significant negative medium-term effects for this group.} \citet{Roger11}, and \citet{Cahuc21} corroborate this conclusion. They demonstrate that (hiring) subsidies for young high school graduates in France do not entail any effects, respectively, on the transition to open-ended contracts, and on callback rates, unless they were targeted at dropouts with certified on-the-job training. Similarly, \citet{Caliendo11} report more positive employment effects of a hiring subsidy for youths with a high school degree than for those less educated. 
	
	However, our findings diverge by revealing a mechanism how "work-first" policies boost long-term earnings. The subsidy enables high school graduates to move from low-wage jobs in non-subsidized sectors, such as the public sector or self-employment, to private sector jobs offering better career prospects but typically rationed due to collectively bargained wage floors. This substitution goes against the existing literature showing crowding out of private sector jobs by public employment \citep{algan2002public, caponi2017effects, fontaine2019labour}. The fact that in our context the private sector jobs are the “good" jobs and the substituted public-sector jobs are the “bad" jobs,  while this is usually the reverse, may explain this opposite finding.  
	
	Third, we are the first to demonstrate empirically that labor market tightness can moderate the effectiveness of hiring subsidies, a prediction made by \cite{Kline13}.\footnote{Previous studies have instead focused on understanding the conditions under which place-based policies can reduce regional inequalities (see \citealp{Glaeser08, Kline14a, Kline14b} for reviews).} We document that the hiring subsidy is a complete deadweight loss close to the border of Luxembourg, a small neighbor country of Belgium. Luxembourg is an economic hub with many attractive employment opportunities which, with the absence of language and legal barriers, attracts an important inflow of cross-border workers.\footnote{Today, 50,000 Belgian residents work in Luxembourg (11\% of the workforce in Luxembourg; \citealp{statec2}).} This important flow of cross-border work increases the labor market  tightness at the other side of the border in Belgium which explains our finding.
	
	This paper is structured as follows. Section \ref{sec:Context} summarizes the institutional setting. The sampling scheme and data are described in Section \ref{sec:Data}. Section \ref{sec:methods} presents the identification strategies and estimation methods. In Section \ref{sec:results}, we present the empirical findings. The last section offers some concluding remarks.

	\section{Institutional Setting}\label{sec:Context}
	
	In December 2009, the Win-Win plan was \emph{unexpectedly} designed and adopted by the Belgian federal government for entry into force on January 1, 2010. It was only on January 18, 2010, that a press release from the Minister of Employment detailed the main features of the plan. This was directly followed by a large advertising campaign on radio, newspapers and Internet.\footnote{As in \citet{Cahuc19}, we use the Google Trends to verify that the introduction of the policy was unexpected. There are no searches for the policy name (“Plan Win-Win” or other variants) until January 2010 but many of them immediately afterwards, induced by the information campaign: see Figure A.1 in Online Appendix A.} The plan involved generous \emph{one-shot} subsidies available for recruitment during two years (2010 and 2011). The hiring subsidies were targeted at the most vulnerable groups of unemployed jobseekers, namely low-educated youths, older workers, and the long-term unemployed. The subsidy was implemented when the economic recovery was underway in Belgium, inducing employment to grow.\footnote{See Figure A.2 in Online Appendix A for a graphical illustration of the evolution of GDP, employment, and unemployment rates.} However, the unemployment rate was still peaking at a high level at the outset of 2010. Youths were particularly hard-hit: In 2009, the unemployment rate of people aged 15-24 rose to 20.4\%, while it was only 7.9\% for the group aged 25-74 \citep{Eurostat23a}.

	In this paper, we evaluate the impact of this reform for what concerns the low-educated youths under 26 years of age (first two rows of Table \ref{tab:win}). The age requirement was verified on the last day before hiring or on the date of the subsidy-eligibility card (see below). Private sector firms recruiting eligible youths benefited from a wage subsidy of about \texteuro1,000 per month for one year (if granted in 2011) or two years (if granted in 2010).\footnote{Specific public sector firms could also benefit from the scheme for the hiring of temporary contractual workers, but this represents a negligible fraction of take-up. In our sample, only 1\% of hiring with a Win-Win subsidy was realized in the public sector.} High school dropouts (graduates) became eligible after only 3 (6) months of registration as jobseekers within the last 4 (9) calendar months. Other jobseekers (post-secondary graduates or aged between 26 and 45) were entitled to a less generous subsidy of \texteuro750 per month during the first year (and \texteuro500 in the second year for recruitments realized in 2010) if they received unemployment benefits, but only to the extent that they had accumulated at least 12 months of unemployment over the last 18 months (but no more than 24 months over the last 36 months - last row of Table \ref{tab:win}).

	The Win-Win subsidy was not awarded automatically. The jobseeker had to prove sufficient unemployment duration to be eligible. To this end, the jobseeker had to fill out a form and request approval from the national Public Unemployment Agency (PUA).\footnote{Eligibility for the Win-Win subsidy did not require jobseekers to receive benefits during the required periods of unemployment. If the unemployed individual was not claiming benefits, the regional Public Employment Service (PES) had to provide proof to the national PUA that this person was officially registered as an unemployed jobseeker during these periods. This complicates the procedure.} The employer then had to draft an appendix to the employment contract, mentioning the subsidy amount that could be deducted directly from the gross salary of the beneficiary worker. The subsidy—referred to as the "work allowance"—was paid directly by the PUA to the worker. This arrangement allowed employers to reduce gross wages below the sectorally agreed-upon or legal minimum, binding only for contractual wages. This feature made it harder for workers to capture subsidy benefits as wage increases. The empirical analysis below finds no evidence of significant passthrough to wages. By contrast, had the subsidy been directly paid to employers without allowing wage reductions below these floors, negotiating pay raises would have been easier.
	
	\begin{table}[htp]
		\caption{Win-Win Hiring Subsidies for Low-Educated Youths and the Long-Term Unemployed Aged Below 45, 2010-2011}
		\label{tab:win}
		\centering\footnotesize\setstretch{0.1}
		\begin{tabular}{@{}cccccc@{}}
			\toprule
			&
			\multicolumn{2}{c}{Registration as unemployed jobseeker} &
			\multicolumn{3}{c}{Wage subsidy} \\ 
			\multicolumn{2}{c}{\phantom{X}} &
			\multicolumn{3}{c}{\phantom{X}} \\ 
			Target &
			\multicolumn{1}{c}{during} &
			in the last &
			\multicolumn{1}{c}{\begin{tabular}[c]{@{}c@{}}Requirements \end{tabular}} &
			\multicolumn{1}{c}{Amount} &
			Duration \\ \midrule
			\begin{tabular}[c]{@{}c@{}}Youth\\ \\ no high school \\ \\ diploma\end{tabular} &
			\multicolumn{1}{c}{\begin{tabular}[c]{@{}c@{}}at least \\\\ \phantom{-} \\\\ 3 months\end{tabular}} &
			4 months &
			\multicolumn{1}{c} {\begin{tabular}[c]{@{}c@{}}Unemployed \\ \\ jobseeker \\ \\ aged below 26\end{tabular}} &
			\multicolumn{1}{c}{\texteuro1,100/month} &
			\begin{tabular}[c]{@{}c@{}}24 months\\ \\ (hiring in 2010)\\ \\ 12 months\\ \\ (hiring in 2011)\end{tabular} \\ \midrule
			\begin{tabular}[c]{@{}c@{}}Youth\\ \\ up to high school \\ \\ diploma\end{tabular} &
			\multicolumn{1}{c}{\begin{tabular}[c]{@{}c@{}}at least \\\\ \phantom{-} \\\\ 6 months\end{tabular}} &
			9 months &
			\multicolumn{1}{c}{\begin{tabular}[c]{@{}c@{}}Unemployed \\ \\ jobseeker \\ \\ aged below 26\end{tabular}} &
			\multicolumn{1}{c}{\texteuro1,000/month} &
			\begin{tabular}[c]{@{}c@{}}24 months\\ \\ (hiring in 2010)\\ \\ 12 months\\ \\ (hiring in 2011)\end{tabular} \\ \midrule
			\begin{tabular}[c]{@{}c@{}}Long-term\\ \\ unemployed \end{tabular} &
			\multicolumn{1}{c}{\begin{tabular}[c]{@{}c@{}}between 12 \\ \\  and 24 months \end{tabular}} & 	\multicolumn{1}{c}{\begin{tabular}[c]{@{}c@{}}	between 18  \\ \\ 	and 36 months  \end{tabular}}	 &
			\multicolumn{1}{c}{\begin{tabular}[c]{@{}c@{}}Insured \\ \\ unemployed \\ \\ jobseeker \end{tabular}} &
			\multicolumn{1}{c}{\begin{tabular}[c]{@{}c@{}}\texteuro750/month\\\\ \phantom{-} \\\\ \phantom{-} \\\\ \texteuro500/month\end{tabular}} &
			\begin{tabular}[c]{@{}c@{}}12 months\\ \\ (hiring in 2010 or 2011)\\ \\ + 16 months\\ \\ (hiring in 2010)\end{tabular} \\ \bottomrule
		\end{tabular}
	\end{table}
	
	If the recruitment was on a part-time basis, the subsidy amount was reduced proportionally. In principle, a firm was not allowed to hire subsidized workers in replacement of other dismissed workers in the same function. The PUA monitored this, but given that only 16 out of the 60,000 examined Win-Win contracts were found to be violating this condition (\citealp{ONEm11}, p. 154), there are doubts about the extent to which non-compliance could be detected.
	
	Insured unemployed jobseekers who were not eligible for the Win-Win subsidy could be eligible for “Activa”, another hiring subsidy that was already in operation before the introduction of the Win-Win plan and which was kept in place. However, Activa was only targeted at long-term unemployment benefit recipients.\footnote{More than 12 months over the last 18 months for those aged under 25 and more than 24 months over the last 36 months for those older than 25.} The subsidy amounted to \texteuro500 per month (for a maximum period of 16 months). Since Activa could not be combined with Win-Win, it was only relevant for the individuals not eligible for Win-Win. Both Win-Win and Activa could be combined with pre-existing deductions for employers' social security contributions.\footnote{These pre-existing measures do not pose a threat to our identification strategy (see Online Appendix C.1).}
	
	The Win-Win plan was the onset of an unprecedented decline in the cost of hiring low-educated youths, demonstrated by its successful adoption. From January 2010 to December 2011, Belgium saw the completion of 101,000 Win-Win employment contracts. Of these, 47\% were allocated to high school dropouts and 23\% to high school graduates, with both groups being under the age of 26. The remaining 30\% were aimed at the long-term unemployed without any age constraints (ONEm11, p. 87).
	
	In the empirical analysis, we exploit the discontinuity in the subsidy amount that the plan induces at age 26.  In our sample of youths registering unemployment in 2010 and taking up a subsidy within one year (see Section \ref{sec:Data}), we observe that slightly below this age, the subsidy amounted to 41\% of wage costs, on average, for high school graduates, and 48\% for dropouts.\footnote{See Online Appendix C.4 for a detailed explanation of how these shares are calculated.} At age 26, these shares drop sharply to 24\% for both group, a decrease of 18 pp for high school graduates and 24 pp for dropouts. The subsidy amounts do not drop to zero, because unemployed aged 26 or older may be eligible for Activa or Win-Win for long-term unemployed if they had accumulated enough months in unemployment. The age-discontinuity at 26 years old therefore results from the sharp decline in the subsidy amount and the more stringent unemployment duration requirement when the low-educated jobseekers reach 26 years. The next sections explain how we exploit this discontinuity to estimate the impact of the Win-Win subsidy on labor market outcomes in the short and long run.

	\section{Data}\label{sec:Data}
	
	The analysis relies on a sample of register data that are collected by various Belgian Social Security institutions and merged into one single database by the Belgian Crossroads Bank for Social Security (CBSS). These data allow reconstructing of individual labor market histories between 2003 and 2017 on a quarterly basis. The sample was originally collected to study cross-border work in Luxembourg from various perspectives. It consists of 125,000 individuals randomly drawn from a stratified population born between December 31, 1972, and December 31, 1990, who lived in Belgium at some point between 2006 and 2017, in a geographical area close to the border with Luxembourg.\footnote{Unemployed and individuals living in municipalities closer to the border of Luxembourg were over-sampled to enhance precision for these groups. The data are appropriately reweighted to take this stratification into account and be representative of the population of interest \citep{Manski77, Cameron05, Albanese19}. Details on the sampling can be found in Online Appendix B.} According to \citet{Eurostat22}, the Belgian Province of Luxembourg was the NUTS-2 region in the EU with the highest incidence of outgoing cross-border workers out of the employed population: 25\% in 2010. Within this area, cross-border work is concentrated in the Grand Duchy of Luxembourg. In 2010, 96\% of all cross-border work in our sample was to Luxembourg, while in the same year 92\% of the total population living in Belgium but working in Luxembourg resided in the sampled areas of the Belgian provinces of Luxembourg or Li\`ege \citep{INAMI10}. The particular selection of individuals living close to the border with Luxembourg allows us to study how labor market tightness across borders can moderate the long-term impact of the reinforcement of the hiring subsidy. 
	
	In this sample, we retain first registrations as unemployed jobseekers at the public employment service (PES) between 2007 and 2012 and follow them with quarterly frequency from the start of their unemployment spell. We cannot determine whether a jobseeker satisfies the unemployment duration requirement for a Win-Win subsidy because we only have information about the unemployment status at the end of the month.\footnote{Because eligibility for the hiring subsidy is based on the number of days of unemployment  over the last 4 or 9 months, the unemployment status at the end of the month cannot precisely capture this eligibility. Hence, some unemployment entries in our sample may benefit from  Win-Win before reaching the eligibility threshold of unemployment duration as defined in our data, i.e. 3 and 6 months, respectively for dropouts and graduates.}  We therefore  only identify \emph{intention-to-treat} effects based on the age requirement for the Win-Win subsidy targeted at youths. 
	
	The benchmark analysis in this paper is conducted on 9,935 young adults with at most a high school degree and aged between 22 and 29. Since Win-Win was abolished by the end of 2011, we retain only unemployment spells that started in 2010 to include individuals who do not lose eligibility for the subsidy within their first unemployment year. To investigate spillover effects on ineligible individuals, we include higher-educated and older youths. Additionally, for the placebo analysis and the differences-in-difference (DiD) estimator implemented as robustness analysis, we include entries into unemployment before and after 2010.
	
	We estimate the impact of the wage subsidy reinforcement on several outcomes, grouped into two categories: exit rates to employment within the first year of unemployment, i.e. in the short run, and employment outcomes extending up to seven years later, i.e. in the long run. We focus at first on private sector employment because most public sector jobs were excluded from the Win-Win subsidy. However, to investigate the mechanisms behind the effect on private sector jobs, we also consider employment types \emph{other} than salaried private sector positions, such as public sector employment and self-employment.

	In the empirical analysis, we control for predetermined explanatory variables such as gender, nationality, household composition, geographical location, work experience, and receiving unemployment benefits, which are measured at entry into unemployment. These covariates are aimed at increasing the precision of the RDD estimator or relaxing the parallel trend assumption of the DiD estimator, as explained in the next section. Descriptive statistics for the explanatory variables and the outcomes are shown in Online Appendix C.2 and C.3.
	
	As Table \ref{descr1} shows, around 20\% of 22-25-year-olds enter a Win-Win job within a year of becoming unemployed. The subsidy take-up does not differ much between the two levels of education. Take-up can occur only if (i) the unemployed satisfies the unemployment duration requirement of 3 or 6 months at the moment of hiring, (ii) the employer applies for the subsidy, and (iii) there is a formal approval by the PUA (see Section \ref{sec:Context}). This explains why it is so low.
	
	For other outcomes, education level matters. The probability of starting a salaried private sector job within one year is 58\% for eligible high school graduates, compared to 44\% for eligible high school dropouts. Similarly, high school graduates worked 12.5 quarters in the private sector over the next seven years, while high school dropouts worked 8.2 quarters. For both groups, this outcome is about 4 quarters higher for those taking the Win-Win subsidy. However, this positive difference favoring Win-Win takers is offset by fewer quarters in other employment forms. The reduction is larger for graduates (2.5 vs. 4.7 quarters) than for dropouts (2.2 vs. 2.5 quarters). This evidence suggests substituting private sector employment for non-private sector employment, for which we provide causal evidence below.
	
	\begin{table}[H]\scriptsize\centering
		\newcolumntype{C}{>{\centering\arraybackslash}X}
		\caption{Selected Descriptive Statistics: Outcomes}\label{descr1}
		\input{descriptives1.tex}
		\caption*{\linespread{0.7}\scriptsize{\textit{Notes:} Mean and standard deviation of the outcome variables. Different groups by column: (1) dropouts aged between 22 and 25 at unemployment entry, (2) dropout Win-Win takers within one year and aged between 22 and 25 at unemployment entry, (3) graduates aged between 22 and 25 at unemployment entry, (4) graduates Win-Win takers within one year and aged between 22 and 25 at unemployment entry. }\normalsize}
	\end{table}
	
	We also compare the explanatory variables between young adults who start a subsidized job and those who do not.\footnote{See Table C.2 in Online Appendix C.3.} In comparison to the latter, the former group tends to more commonly be of Belgian nationality, live alone, receive unemployment benefits at registration, have some previous work experience, and have previously benefited from activation policies. This means that the subsidized group is positively selected, and hence, the above descriptive statistics of outcomes cannot be given a causal interpretation.

	\section{Identification Strategies and Estimation Methods}\label{sec:methods}
	
	\subsection{Regression Discontinuity Design}\label{sec:RDD}
	To estimate the causal impact of the wage subsidy reinforcement resulting from the Win-Win plan on the employment trajectories of unemployment entrants, we exploit two eligibility conditions: age and calendar time.\footnote{As mentioned in Section \ref{sec:Data}, unemployment duration's imperfect measurement in the data prevents using eligibility thresholds based on duration for analysis. Consequently, we identify only intention-to-treat effects due to the satisfaction of the age-eligibility condition.} Indeed, only jobseekers younger than 26 and recruitments in 2010 or 2011 are potentially eligible for the most favourable Win-Win subsidies. Our benchmark analysis relies on a regression discontinuity design (RDD) estimator that exploits the age eligibility cutoff at 26 for the unemployed registering in 2010. As mentioned in Section \ref{sec:Context}, workers older than 26 can be eligible for other \emph{lower} hiring subsidies, such as \emph{Activa} or Win-Win targeted at the long-term unemployed. This means that the counterfactual we estimate  is not the absence of eligibility but the age-eligibility for less generous hiring subsidies.

	Because the age cutoff is not determined at a fixed point in time, but at hiring, we cannot implement a standard RDD using age at unemployment registration as the forcing variable. Youths slightly younger than 26 at entry into unemployment will indeed gradually, along the unemployment spell, age out of eligibility for the higher subsidy for youths in the Win-Win plan. We address this issue as in \cite{Gerard21} and ignore observations of individuals aged between 25 and 26 at the start of the unemployment spell. By doing so, we ensure that all individuals younger than 25 years do not age out of their eligibility for a higher subsidy before the end of the first year of unemployment. By dropping these observations, we create a ``hole" to the left of the age cutoff of 26, which we fill by the prediction of a linear spline estimated using data points to the left of this ``hole". 
	
	This approach resembles the so-called \emph{donut} RDD \citep{Barreca16}, which in the literature is used to solve another identification problem, namely \emph{manipulation} around the threshold. In the latter problem the extrapolation is required at both sides of the discontinuity threshold, whereas here the extrapolation is implemented only on the left-hand side. This extrapolation enables us to identify the intention-to-treat (ITT) effects resulting from meeting the age and time eligibility conditions for at least one year after entering unemployment.\footnote{In Online Appendix F, we discuss how to estimate the local average treatment effect (LATE) and argue that it represents a less relevant  policy parameter in the context of hiring subsidies.} Details on the formal implementation of the estimator can be found in Online Appendix D, while the validation analysis, such as the balancing and density tests, is shown in Subsection \ref{sec:sens}.
	
	In empirical applications implementing an RDD estimator, it has become standard practice to rely on the optimal bandwidth selector of \citet{Calonico14}. However, this selector aims to find the \emph{local} non-parametric estimator that minimizes the mean square error at cutoff. Since we cannot use the observations in the donut to the left of the cutoff, this selector is not well defined. We therefore set the bandwidth ad hoc, at three years for each side of the discontinuity (outside the donut). In Section \ref{sec:robustness}, we then test the sensitivity of the results to wider or narrower bandwidths, besides implementing other sensitivity analyses. This shows that the results are robust. Finally, to take into account that the running variable, age, is grouped in monthly intervals, we cluster the standard errors by age in months \citep{Lee08}. In the benchmark analysis, this defines 72 clusters. The units are reweighted by using the triangular kernel and the sampling weights to make inference on the population.

	\subsection{Difference-in-Differences}\label{sec:DiD1}
	
	By applying the donut RDD the treatment effects are no longer completely non-parametrically identified. We, therefore,  check whether the main RDD estimates are robust to a difference-in-differences (DiD) estimation strategy.
	
	In the DiD design we contrast the evolution of the outcomes, each measured over a fixed horizon since entry into unemployment, between a treated group (aged between 24 and 25 at entry) and a control group (aged between 26 and 27). We consider two entry periods: 2008 and 2010. These are, respectively, the control period and the intervention period.\footnote{We do not use the unemployed entering in 2009 since they quickly enter the treatment period in 2010.} The treatment applies only to the treated group that entered unemployment in 2010. Those aged 24-25 at entry and registering into unemployment in 2010 are eligible for the Win-Win subsidy for at least one year of unemployment (up to 2 years) because these individuals are younger than 26 during the full period that the Win-Win subsidy was in place (2010-11).\footnote{We do not consider those aged between 25 and 26 at entry into unemployed or unemployment registrations in 2011 as they are potentially eligible for less than one year.} By contrast, the members of the treated group entering in 2008 are never eligible for the subsidy because they are older than 26 by 2010. Individuals in the control group are never eligible for the subsidy because they are already older than 26 from entry into unemployment.
	
	The main identifying assumption is that the counterfactual outcomes in the absence of treatment of the treated and the control group follow parallel trends. A potential threat is that the business cycle has a steeper age gradient on employment outcomes for youths than for prime-aged workers, thereby invalidating the parallel trends assumption. This threat is less forceful here because the age differences between the treated [24-25] and control group [26-27] are only minor \citep[See e.g.,][]{Meyer95}. In Section \ref{sec:estimatorDD}, we present a series of validation test that support the credibility of parallel trends. The treatment effect is then estimated by  relaxing the parallel trend assumption to hold only conditional on our predetermined control variables. We do this by employing the doubly robust DiD estimator, as proposed by \citet{SANTANNA20}. Further details can be found in Online Appendix E.

	\section{Empirical Findings}\label{sec:results}
	
	This section reports the empirical results of our analysis in three main sections. First, the findings are discussed based on graphical evidence resulting from the (donut) RDD, which is our benchmark identification strategy. The econometric estimates and associated statistics underlying these graphs are reported in Online Appendix H. Second, we demonstrate that our empirical findings are robust to various validation and sensitivity tests, and that the DiD estimator replicates the estimates. Finally, we present the results of a cost-benefit analysis following the marginal value of public funds framework proposed by \citet{Hendren20}.   
	
	\subsection{Main results}\label{sec:main}
	We start by documenting the incidence of the Win-Win subsidy on wages and employment in the private sector. We show that this reinforcement significantly enhances the transition to employment within one year of entry into unemployment for both high school dropouts and graduates, and that there is little incidence on wages. The effect on private sector employment and unconditional earnings persists beyond the expiration of the subsidy for high school graduates only. In a next subsection, we investigate where the persistent effects for graduates come from. We show that the subsidy reinforcement induces substitution effects between private employment and other precarious forms of employment in unsubsidized sectors and discuss the mechanism that brings this about. Next, we provide evidence that labor market tightness leveraged by the geographic proximity of the economic hub of to Grand Duchy of Luxembourg moderates the treatment effects for graduates. We also investigated whether subsidized employment generated negative spillovers on ineligible older workers. In line with the existing evidence \citep{Blundell04, Kangasharju07, Pallais14, Webb16, Cahuc19, Saez21}, we do not find evidence of such  spillovers. For the latter analysis, the interested readers are referred to Online Appendix E.2.
	
	\subsubsection{Incidence of the Hiring Subsidy on Employment and Wages in the Private Sector}\label{secpriv}
	\emph{The Reinforcement of the Subsidy Conditional on Hiring} 
	\medskip

	We estimate the impact of meeting the age-eligibility criteria for the Win-Win plan on the average subsidy amount \emph{among hired jobseekers}. Two sources of non-compliance must be considered when analyzing this outcome.
	
	First, there is non-compliance because long-term unemployed individuals older than 26 are also eligible for a lower subsidy than those younger than 26. By applying the RDD estimator to the average full-time equivalent subsidy \emph{conditional on subsidy uptake}, we estimate that jobseekers younger than 26 (to the left of the age cutoff) are entitled to a subsidy of \texteuro1,056 per month when entering a subsidized job (the average of \texteuro1,100 per month for high school dropouts and \texteuro1,000 per month for graduates), while older jobseekers (to the right of the cutoff) receive \texteuro543 per month (the average of \texteuro750 per month for those eligible for the Win-Win subsidy for long-term unemployed and \texteuro500 per month for those eligible for the Activa plan). Thus, the additional reinforcement due to age eligibility is \texteuro513 per month.\footnote{See Figure F.1 in Online Appendix F.} Therefore, the comparison of individuals on both sides of the cutoff measures the effect of this reinforcement at the age threshold, rather than the total effect of the Win-Win subsidy for low-educated youths.
	
	Second, irrespective of age, not all \emph{hired} unemployed are eligible for a subsidy because for some the unemployment duration requirements are not met and for others firms do not comply with the administrative formalities, either because they are not informed or because they find the administrative hurdles too costly.\footnote{Our data do not allow to distinguish between these three reasons.} The share of subsidized hires among all hires (i.e., the \emph{``attention rate"}) is 36\% to the left of the age cutoff at 26 and 14\% to the right.\footnote{Online Appendix C.4 explains how these attention rates are estimated.} The attention rate is lower to the right than to the left of the cutoff because the subsidy is lower and the eligibility criteria are stricter for youths aged 26 or more.\footnote{Youths aged 26 or more must be at least one year unemployed to be eligible for the subsidy, while below 26 this requirement drops to 3 and 6 months, respectively for high school dropouts and graduates (see Table \ref{tab:win}).} This partial eligibility reduces the expected subsidy that firms receive when hiring these unemployed youths. The average  subsidy conditional on hiring on the left and right of the cutoff is 0.36x\texteuro1,056 = \texteuro380/month and 0.14x\texteuro543 = \texteuro76/month, respectively. The difference (\texteuro304 = \texteuro380 - \texteuro76) is the average subsidy reinforcement that firms receive for hiring workers slightly below age 26.  
	
\begin{figure}[H]
		\centering
		\caption{Discontinuity of Employment Outcomes in the Private Sector Within One Year}\label{all:alltake}
		\begin{subfigure}{0.496\textwidth}
			\centering
			\caption{\fontsize{10}{12}\selectfont Subsidy Amount Conditional on Employment}
			\includegraphics[ width=0.8\textwidth]{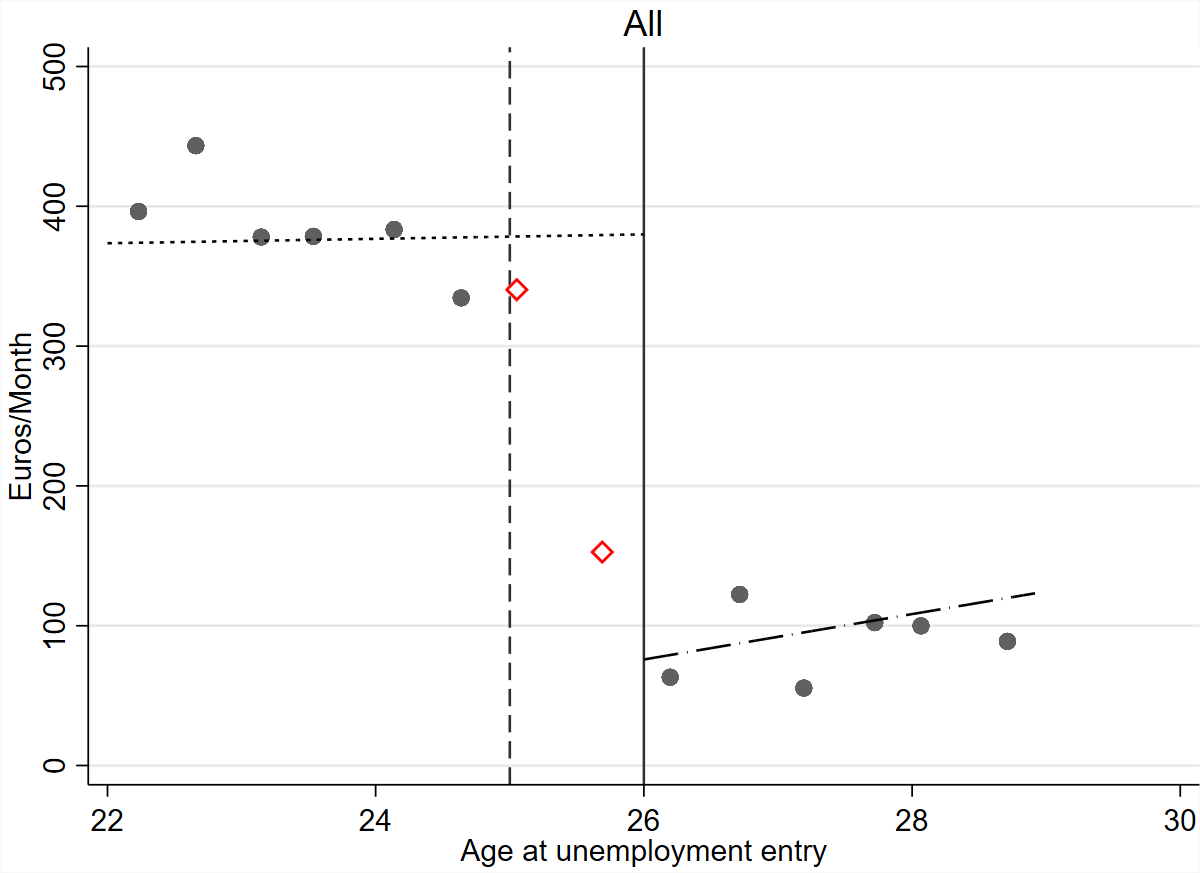}
		\end{subfigure}	
		\begin{subfigure}{0.496\textwidth}
			\centering
			\caption{\fontsize{10}{12}\selectfont Full-time Gross Wage Conditional on Employment} 
			\includegraphics[ width=0.8\textwidth]{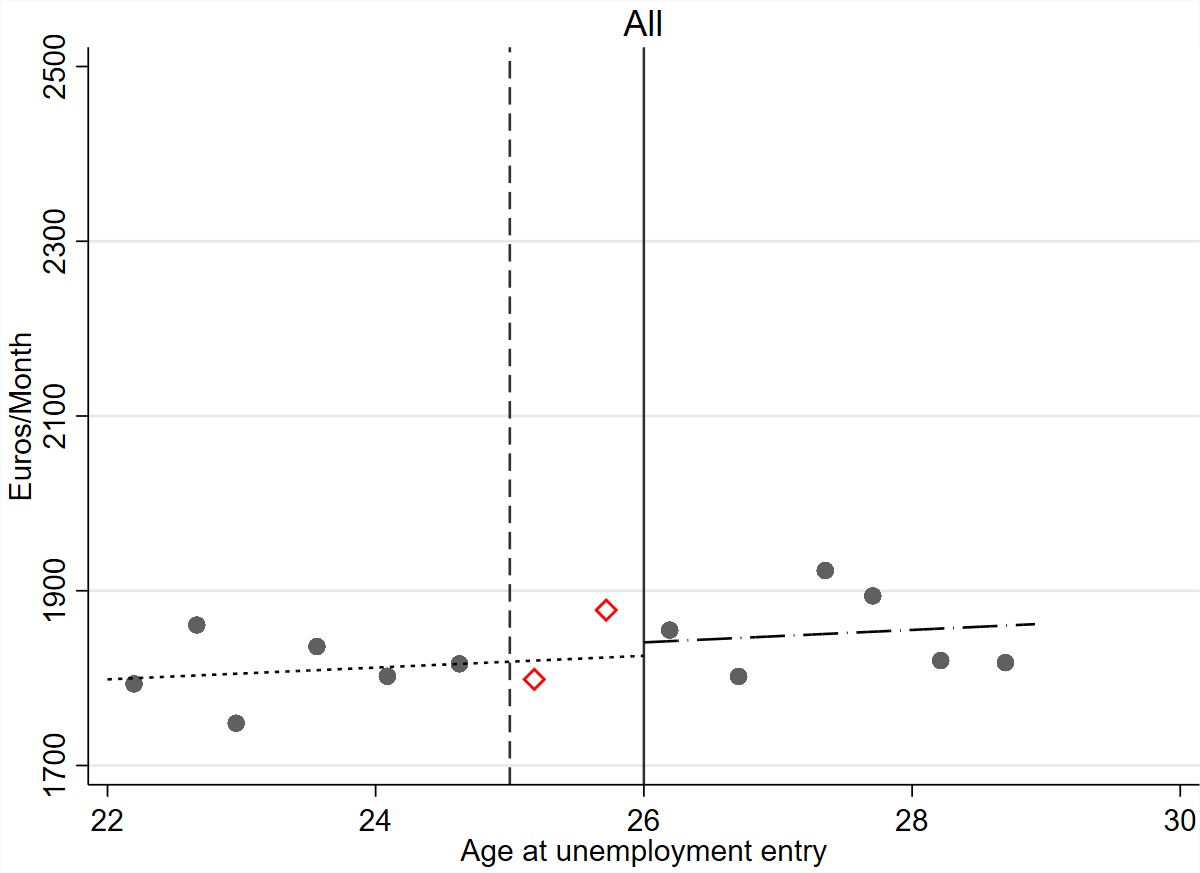}
		\end{subfigure}	
		\begin{subfigure}{0.496\textwidth}
			\centering
			\caption{\fontsize{10}{12}\selectfont Full-time Wage Costs Conditional on Employment}
			\includegraphics[ width=0.8\textwidth]{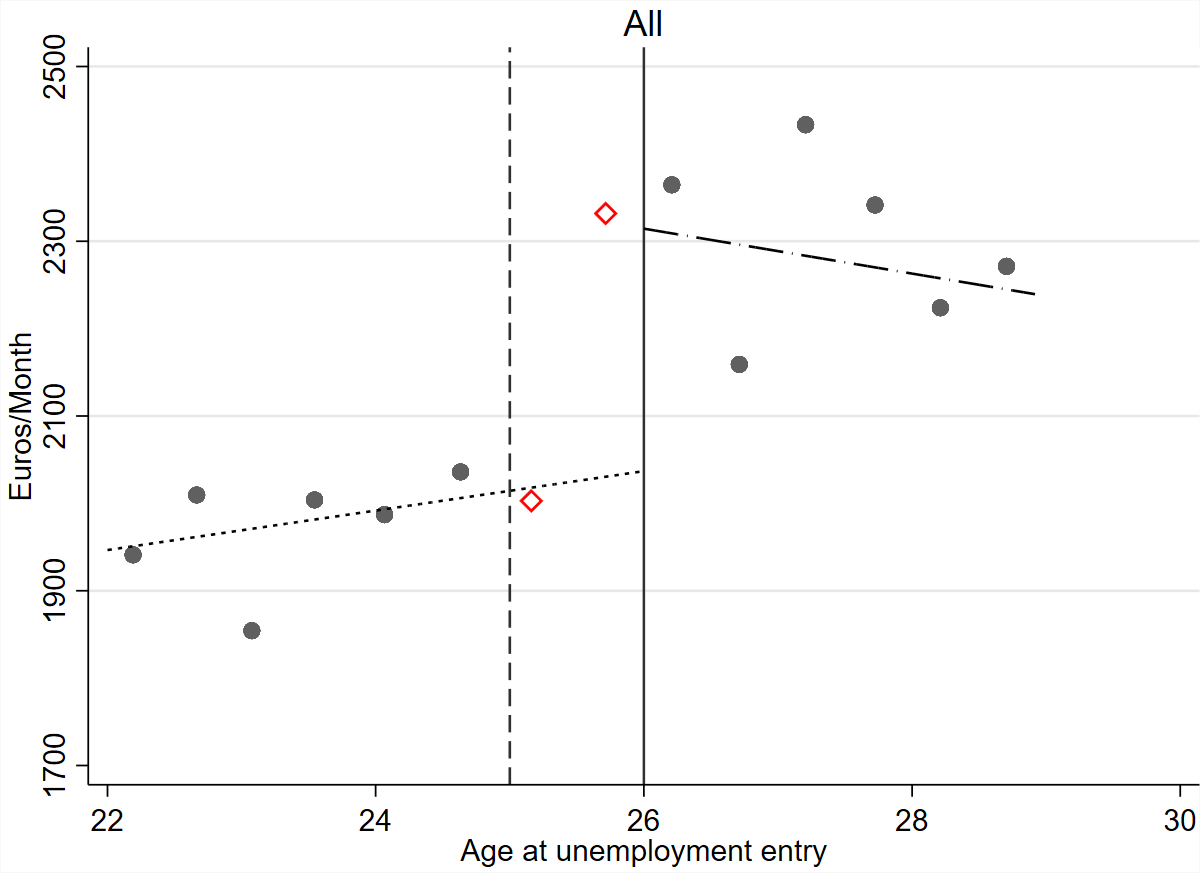}
		\end{subfigure}	
		\begin{subfigure}{0.496\textwidth}
			\centering
			\caption{\fontsize{10}{12}\selectfont Cumulative Transition Rate to Employment}
			\includegraphics[ width=0.8\textwidth]{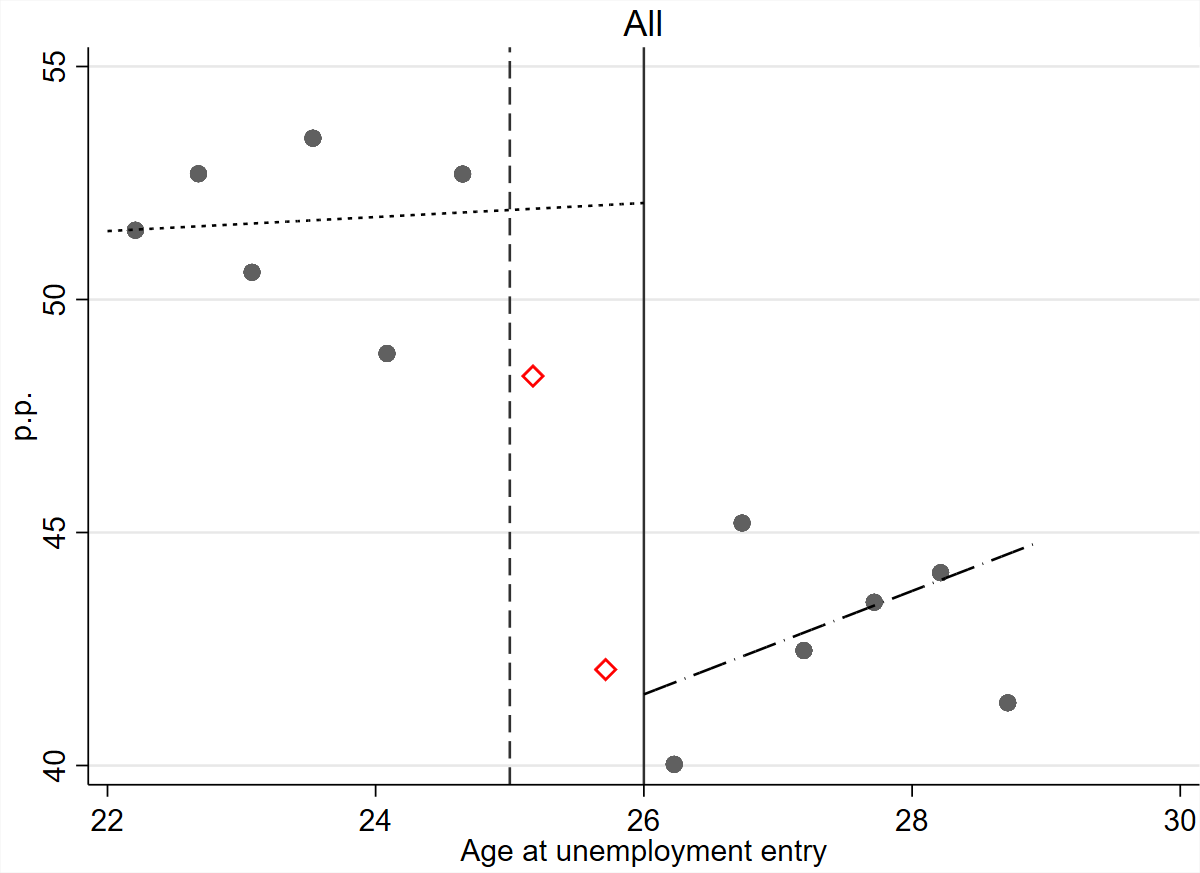}
		\end{subfigure}	
		\caption*{\footnotesize{Note: Donut RDD estimate on the inflow sample of youths entering unemployment in 2010, using age at unemployment entry as the forcing variable with a cutoff at 26.   The outcome is (a) the amount of received subsidy (in full-time equivalent) conditional on hiring in the private sector,  (b) the full-time gross wage conditional on hiring in the private sector, (c) the full-time wage cost conditional on hiring   in the private sector and (d) the cumulative transition rate to private sector employment within one year after unemployment entry, which is plotted over six age-quantile-spaced bins on each side of the donut. The two local linear splines are estimated on the reweighted observations by using the triangular kernel and the sampling weights but by removing the units within the donut. We control for the set of control variables shown in Table C.2 in Online Appendix C.3.  Standard errors are clustered at the age level. The effect estimated by the donut RDD estimator at 26 years old is (a) \texteuro304 [185; 423] with a p-value of 0.000 and N = 3,958,  (b) --\texteuro15   [--104; 74] with a p-value of 0.733 and N = 3,958, (c) --\texteuro277   [--438; --117] with a p-value of 0.001 and N = 3,958 and (d) +10.5 pp [3.0; 18.1] with a p-value of 0.007 and N = 8,560. 
		}}
	\end{figure}

	Panel (a) of Figure \ref{all:alltake} shows the average subsidy received \emph{conditional on being hired} within one year after entering unemployment, binned by age at entry into unemployment grouped into six-month intervals. The linear splines and the discontinuity at the age cutoff are estimated by the donut RDD procedure described in Section \ref{sec:RDD}. The donut excludes the red diamonds reporting the expected subsidy for youths aged between 25 and 26 at entry into unemployment. These diamonds are excluded because the corresponding youths are only eligible for the Win-Win subsidy part of the year, i.e. until their $26^{th}$ anniversary. The linear splines to the left and right cut the age cutoff at the expected subsidy levels mentioned above, i.e. \texteuro380 and \texteuro76. The difference (\texteuro304/month)  is highly significant (with a p-value close to 0\%), which demonstrates that the treatment is strong. This reinforcement represents 12.7\% of the full-time monthly wage cost in the absence of the subsidy reinforcement, as estimated by the wage costs to the right of the cutoff (\texteuro2,392) in panel (c).\footnote{Contrary to the expected subsidy for which the generosity is fixed, this wage cost is possibly a biased estimate of the mentioned counterfactual though we show in the next subsection that this bias is small.}  This share does not differ by educational attainment.\footnote{For high school dropouts and graduates the differential monthly subsidy is \texteuro297 and \texteuro323, respectively (see Figure A.3 in Online Appendix A).}
	
	\noindent\emph{The Incidence of the Hiring Subsidy on Wages} 
	\medskip
	\\
	The effect of the hiring subsidy on employment depends on the extent to which workers can capture part of the subsidy through higher wages. Panel (b) of Figure \ref{all:alltake} displays the RDD for the gross before-tax wage measured by dividing the FTE gross wage of youths hired within the first year since entry into unemployment by the number of quarters in employment. The gross wage does not display any discontinuity at the age cutoff of 26. Even if the subsidy is borne by the worker, there is a complete passthrough to the employer. Nevertheless, it is difficult to directly infer the effect on this outcome as it suffers from a ``double selection problem'', which may bias the estimates \citep{heckman1974shadow}. We obtain some insight into the extent to which this bias matters by comparing the RDD on the wage costs in panel (c) to the RDD on the average  subsidy conditional on hiring reported in panel (a) of Figure \ref{all:alltake}.

	Panel (c) displays the RDD on the FTE wage costs including employer's social insurance contributions \emph{net} of wage subsidies. The point estimate is minus \texteuro277/month. This amount of cost savings at the cutoff is very close to the average subsidy reinforcement of \texteuro304/month estimated in panel (a) (which is by construction not affected by the selection bias mentioned above as the subsidy amounts are exogenously fixed). This suggests that the selection bias is only minor and therefore that the passthrough of the subsidy to gross before-tax wage, if any, is not large. This may be very much linked to the fact that the subsidy allows employers to reduce the wage below the legal or sectoral minimum wage by the amount of the subsidy. This differs from the usual implementation scheme in which the employer receives the subsidy but cannot reduce the wage below these floors (see Section \ref{sec:Context}).
	
	An explanation for this limited incidence on the workers' wages is that pay scales in Belgium are negotiated in collective agreements covering all workers in a sector, and firm specific wage top-ups are not common. It is therefore difficult to negotiate a different wage for hired workers in a specific age range than for incumbents. However, we cannot exclude that there is  wage passthrough shared by all workers in firms. \citet{Saez19}, for example, find no incidence on the net of payroll tax wages due to a cut in employer's social contributions for employed youths in Sweden. However, they do observe  positive effects on the wages of incumbents. We do not have the data to check this. Nevertheless, even if such a passthrough exists, it cannot be important because hiring subsidies are targeted at a much smaller group in the firm than a payroll tax cut for all incumbent young workers. To the extent that labor demand is elastic, we expect a positive impact on the hiring rate.
	
	\par\vspace{0.5\baselineskip}
	
	\emph{The Effect on Hiring and the Labor Demand Elasticity in the Short Run}
	
	\noindent Panel (d) of Figure \ref{all:alltake} shows the donut RDD estimates for the transition rate to private sector employment one year after unemployment in 2010 are 10.5 pp (significant at the 0.7\% level), a 25\% increase over the counterfactual. This demonstrates that the subsidy reinforcement for youths positively impacted this rate.
	
	When we split the sample by education, binned data points become more noisy and estimates, therefore, less precise.\footnote{See Figure A.4 in Online Appendix A.} For high school graduates and dropouts, the point estimates are 8 pp and 13 pp (p-values, 0.25 and 0.02), which correspond to proportional increases of 15\% and 40\% relative to the counterfactual. However, the finding that the subsidy reinforcement does not significantly affect the hiring rate for high school graduates is not robust. We also report the evolution in the RDD estimates of these cumulative transition rates by educational attainment from one to six quarters after entry into unemployment.\footnote{See Figure A.5 in Online Appendix A.}  This shows that the estimate at four quarters spikes downward for graduates (and upward for dropouts). There is therefore no clear evidence that the effect in pp differs between the two education groups. However, relative to the counterfactual, this implies a bigger effect for dropouts than graduates.
	
	Next, we combine the information of panels (a) and (d) of Figure \ref{all:alltake} to obtain an estimate of the employment elasticity with respect to change in wage costs induced by the incremental hiring subsidy, i.e. the labor demand elasticity to the extent that there is no passthrough to employees' wages. When we group the two education groups, this elasticity is estimated by the ratio of the proportional increase in the hiring rate (i.e. 25.4\%) to the proportional decline in wage costs (i.e. $-12.7$\%),\footnote{This is computed by dividing the subsidy amount effect in panel (a) of Figure \ref{all:alltake} (\texteuro304/month) by the predicted wage costs at the cutoff in panel (c) (\texteuro2,392/month). In Online Appendix C.5, we discuss how adjusting for minor bias in the predicted wage costs has negligible effects on elasticity.} and is, hence, equal to $-2.0$.\footnote{For dropouts it is larger ($40.4\%/(-13.0)\%=-3.1$) than for the graduates ($14.6\%/(-13.1)\%=-1.1$), but in view of the lack of robustness of the effects by educational attainment, these point estimates are less reliable.}  The estimate of this elasticity is in the range reported in several studies evaluating the impact of hiring subsidies targeted at long-term unemployed: $-2.5$ and $-2.2$ in the studies of \cite{Ciani19} and \cite{pasquini19} for Italy, and $-1.0$ for a hiring subsidy for prime-aged long-term unemployed in Belgium \citep{Desiere22}. By contrast, in Sweden the corresponding elasticities are reported to be lower. They range between $-0.2$ and $-0.6$ \citep{Sjogren15}.\footnote{These elasticities are not directly comparable to the standard wage elasticities of labor demand \citep{Lichter15}, or the higher one for hiring subsidies reported by \cite{Cahuc19}. The latter elasticities measure how an increase of wage costs affects employment in firms, while the elasticity that we report here measure how this affects the probability that a worker is hired.}
	
	These elasticities only provide a sense of the magnitude of the subsidy effects on hiring in the short run. In the next section we investigate whether these effects persist beyond the end of the subsidy period in the long run.	
	\medskip
	
	\noindent\emph{Cumulative Effects in the Long Run} 
	\medskip
	
\begin{figure}[H]
		\centering
		\caption{Evolution of the RDD Effect on the Cumulative Number of Quarters in Subsidized Private-Sector Employment}\label{fg:ATTsub}
		\includegraphics[width=0.6\textwidth]{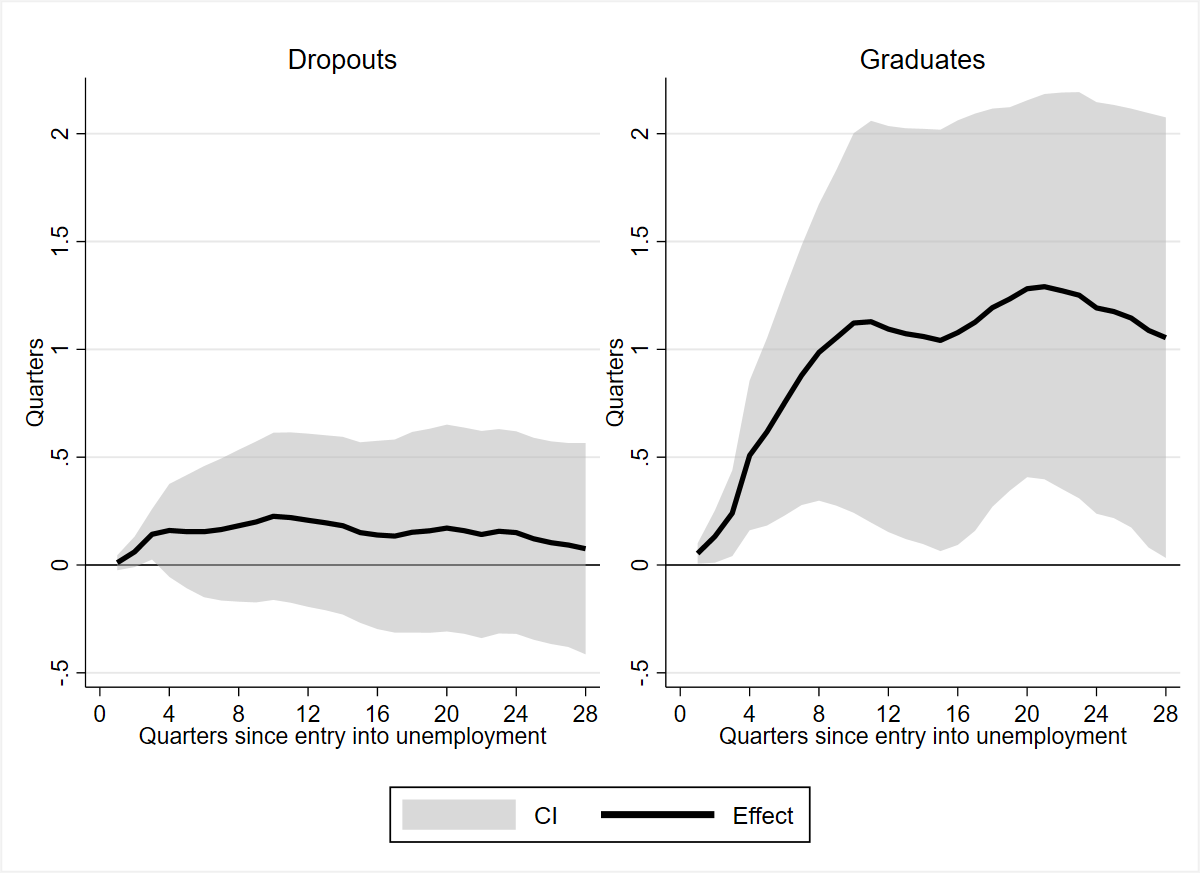}
		\caption*{\footnotesize{Note: Donut RDD estimates on the inflow sample of youths entering unemployment in 2010, using age at unemployment entry as the forcing variable with a cutoff at 26. Evolution of the RDD effect and confidence interval (CI) for the unconditional cumulative number of quarters in subsidized private sector employment by schooling level: dropouts (left) vs. graduates (right). The RDD estimator is implemented for each quarter after entry into unemployment until 7 years later. The donut RDD estimator removes observations for those aged between 25 and 26 at entry into unemployment and retains units aged between 22 and 25 (left of the cutoff) and 26 to 29 (right of the cutoff). The two local linear splines are estimated on the reweighted observations by using the triangular kernel and the sampling weights but removing the units within the donut. We control for the set of control variables shown in Table C.2 in Online Appendix C.3.  Standard errors are clustered at the age level. For dropouts (graduates), the effect at quarter 11 is +0.2 quarters [--0.2; 0.6] with a p-value of 0.279 and N = 4,176 (+1.1 quarters [0.2; 2.0], p-value 0.018 and N = 4,384).}}
	\end{figure}

	\noindent	Figure \ref{fg:ATTsub} shows, by educational attainment, the evolution of the cumulative number of quarters in \emph{subsidized} employment from entry into unemployment in 2010 until seven years later.\footnote{This is an unconditional outcome which considers all entrants into unemployment in 2010.} This number should attain a maximum around quarter 11 when the Win-Win subsidy expires for all unemployment entries.\footnote{The benefits expire between 4 and 11 quarters after entry into unemployment, depending on the calendar year in which the subsidy is claimed and when subsidized employment begins following the start of unemployment.}  For high school graduates, the cumulative effect fluctuates after 11 quarters around the same level of slightly more than one quarter per person in this group. A general observation that also applies to the next graphs is that these long-term effects are estimated with considerable imprecision so we cannot say much about the quantitative effect sizes. On the other hand, when we estimate the same model by DiD on the individuals aged between 24 and 25, the point estimates differ hardly.\footnote{See Figure A.6 in Online Appendix A.}

	The most striking observation is that around the expiration of the subsidy, the cumulative effect of the Win-Win Plan on the average number of quarters in subsidized employment is five times smaller for high school dropouts than for high school graduates. This result implies that subsidized employment tends to be of much shorter duration for dropouts and already suggests that the employment effect in the long run must be small for this group.
	
\begin{figure}[H]
		\centering
		\caption{Discontinuity at Age 26 for the Cumulative Number of Quarters in Private Sector Employment Seven Years after Entry into Unemployment}\label{fg:EMP28}
		\includegraphics[ width=0.85\textwidth]{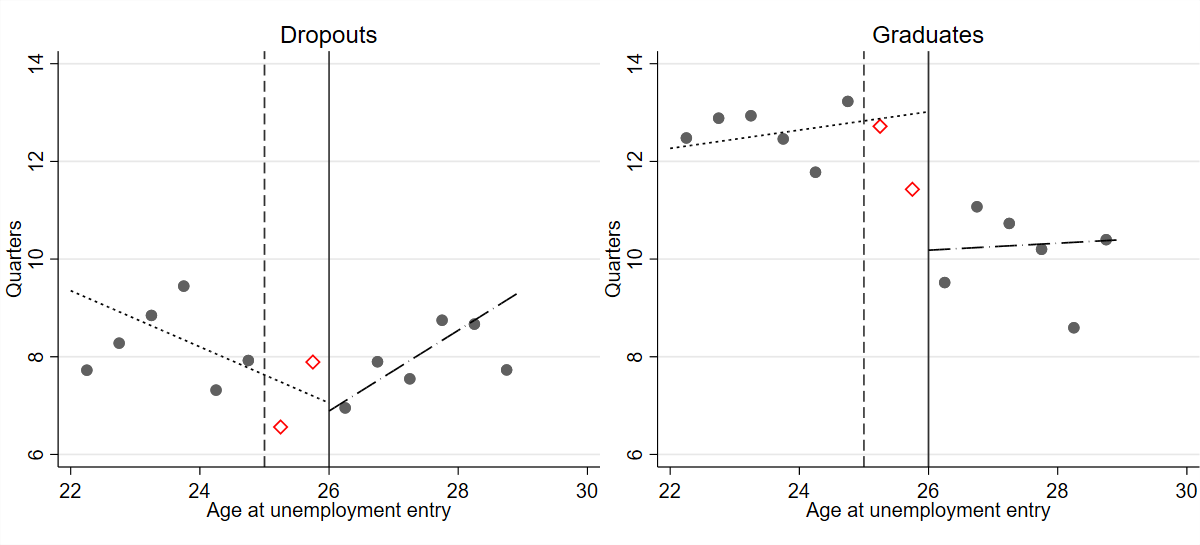}
		\caption*{\footnotesize{Note: Donut RDD estimate on the inflow sample of youths entering unemployment in 2010, using age at unemployment entry as the forcing variable with a cutoff at 26.   The outcome is the cumulative number of quarters in private sector employment seven years after entry into unemployment by schooling level (dropouts vs. graduates), which is plotted over six age-quantile-spaced bins on each side of the donut. The two local linear splines are estimated on the reweighted observations by using the triangular kernel and the sampling weights but removing the units within the donut. We control for the set of control variables shown in Table C.2 in Online Appendix C.3.  Standard errors are clustered at the age level. The effect estimated by the donut RDD estimator at 26 years of age is +0.2 quarters [--2.4; 2.7] (p-value 0.897) and N = 4,176 for dropouts, while for graduates it is +2.8 quarters [0.7; 5.0] (p-value 0.011) and N = 4,384.}}
	\end{figure}

	To test this hypothesis, we estimate the effect of the reinforcement by schooling level on the accumulated number of quarters in private sector employment seven years after entering unemployment. From Figure \ref{fg:EMP28}, we can deduce that the Win-Win subsidy did not affect the time spent in private sector employment for dropouts at such a long-time horizon. During the first 11 quarters, the average number of quarters in employment do slightly increase, but this effect is not statistically significant. Moreover, after the expiration of the subsidy, the estimated effect gradually drops to zero.\footnote{In Figure A.7 in Online Appendix A, we present the evolution of the RDD effects.}  This is evidence that for dropouts, the hiring subsidy only accelerates the transition to short-term jobs and does not generate any persistent effect on employment.

	In contrast, the panel to the right of Figure \ref{fg:EMP28} shows that the Win-Win subsidy increases the average number of quarters in private employment up to 2.8 quarters seven years after entry into unemployment. This is an increase of 28\% relative to the counterfactual of less favorable hiring subsidy conditions. This effect continues to grow beyond the end of the subsidy period and is statistically significant at the 5\% level from quarter 12 onwards. We also estimate that the gains are in terms of FTE employment. After seven years, about three FTE quarters of employment are gained, on average (+26\%).\footnote{See Figure A.8 in Online Appendix A.} The effects on gross wage earnings (assigning zero earnings to those who are not employed in the salaried private sector) follow a similar pattern to those on the number of quarters spent in private sector employment: no effect for dropouts, and for graduates, a steady increase until \texteuro14,600 after 5.5 years, beyond which the effect stabilizes.\footnote{See Figure A.9 in Online Appendix A.} Relative to the counterfactual, the increase after 7 years is 29\%, which is only slightly larger than the effect in full-time equivalent quarters (26\%). Combining these two pieces of evidence suggests that in the long run, the subsidy does not have a significant impact on the wage rate. This is corroborated by the effect estimates on the average full-time equivalent (FTE) gross wage over time, which indicate an effect that is consistently close to zero.\footnote{See Figures A.10 and A.11 in Online Appendix A.} However, these wage effects are conditional on employment and therefore subject to the selection bias mentioned above.

	These findings suggest that "work-first" policies can be effective in the long run for high school graduates. The fact that the subsidy reinforcement does not entail any effect for high school dropouts is in line with the literature that argues that a minimum skill level is required for initiating a process of human capital accumulation on the job (see, for example, \citealp{Card05, Autor10, Cahuc21}). However,  the mechanism is more complicated. In the next section we demonstrate that the job creation in the private sector for high school graduates comes at the expense of low-paid employment in other sectors.
	
	\subsubsection{Mechanisms}\label{sec:mech}

\begin{figure}[H]
		\centering
		\caption{Donut RDD Plot of the Effect on the Cumulative Number of Quarters in Non-Private Sector Employment 7 Years After Entry into Unemployment}\label{afg:OTHER28}
		\includegraphics[ width=0.85\textwidth]{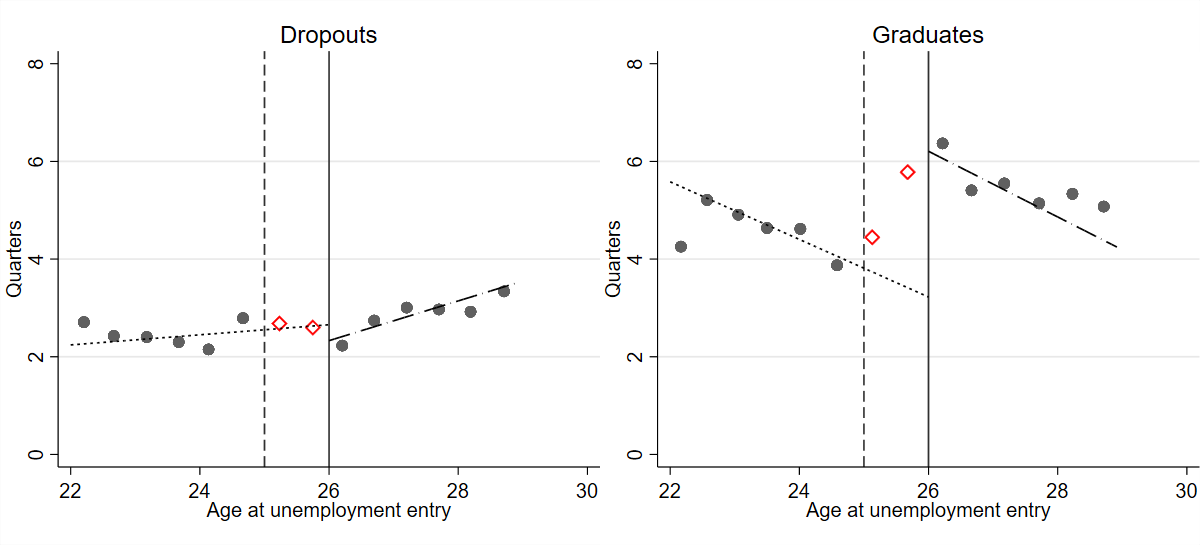}
		\caption*{\footnotesize{Note: Donut RDD estimate on the inflow sample of youths entering unemployment in 2010, using age at unemployment entry as the forcing variable with a cutoff at 26.   The outcome is the cumulative number of quarters in non-private sector employment 7 years after entry into unemployment by schooling level (dropouts vs. graduates), which is plotted over six age-quantile-spaced bins on each side of the donut. The two local linear splines are estimated on the reweighted observations by using the triangular kernel and the sampling weights but removing the units within the donut. We control for the set of control variables shown in Table C.2 in Online Appendix C.3.  Standard errors are clustered at the age level. The effect estimated by the donut RDD estimator at 26 years of age is +0.5 quarters [--0.6; 1.5] (p-value 0.375) and N = 4,176 for dropouts, while for the graduates it is --2.6 quarters [--4.7; --0.6] with a (p-value 0.012) and N = 4,384. }}
	\end{figure}

	Figure \ref{afg:OTHER28} displays the donut RDD effect at the age cutoff of 26 years on the cumulative number of quarters in non-private sector employment, i.e., public sector employment and self-employment, seven years after entry into unemployment. For high school graduates, one can see that the plot displays opposite effects to the ones showed in Figure \ref{fg:EMP28} for private employment. The point estimate is significantly negative and equal to –2.6 quarters. For dropouts it is slightly positive (+0.5) but non significantly. The overall effect on total employment is therefore very close to zero after 7 years for both dropouts and graduates.\footnote{See Figures A.12 in Online Appendix A.}

	For high school graduates, 80\% of the negative effect on non-private sector employment comes from reduced public employment (–2.1 quarters) and  20\% from self-employment (–0.5 quarters).\footnote{See Figures A.13-A.14 in Online Appendix A.} An important observation is that this substitution only appears gradually after a couple of years. If we backtrack on cumulative transition rates in the short run (10.5 pp - see  Subsection \ref{secpriv}), the enhanced job finding rate in the private sector comes from fewer youths remaining unemployed. No negative effect on the transition to self- or public sector employment is observed in the first year after entry into unemployment.\footnote{See Figures A.15 in Online Appendix A.}

	\begin{figure}[H]
		\centering
		\caption{Discontinuity at Age 26 for the Cumulative Labor Income Tax Revenue Seven Years after Entry into Unemployment}\label{fg:tax28un}
		\includegraphics[ width=0.85\textwidth]{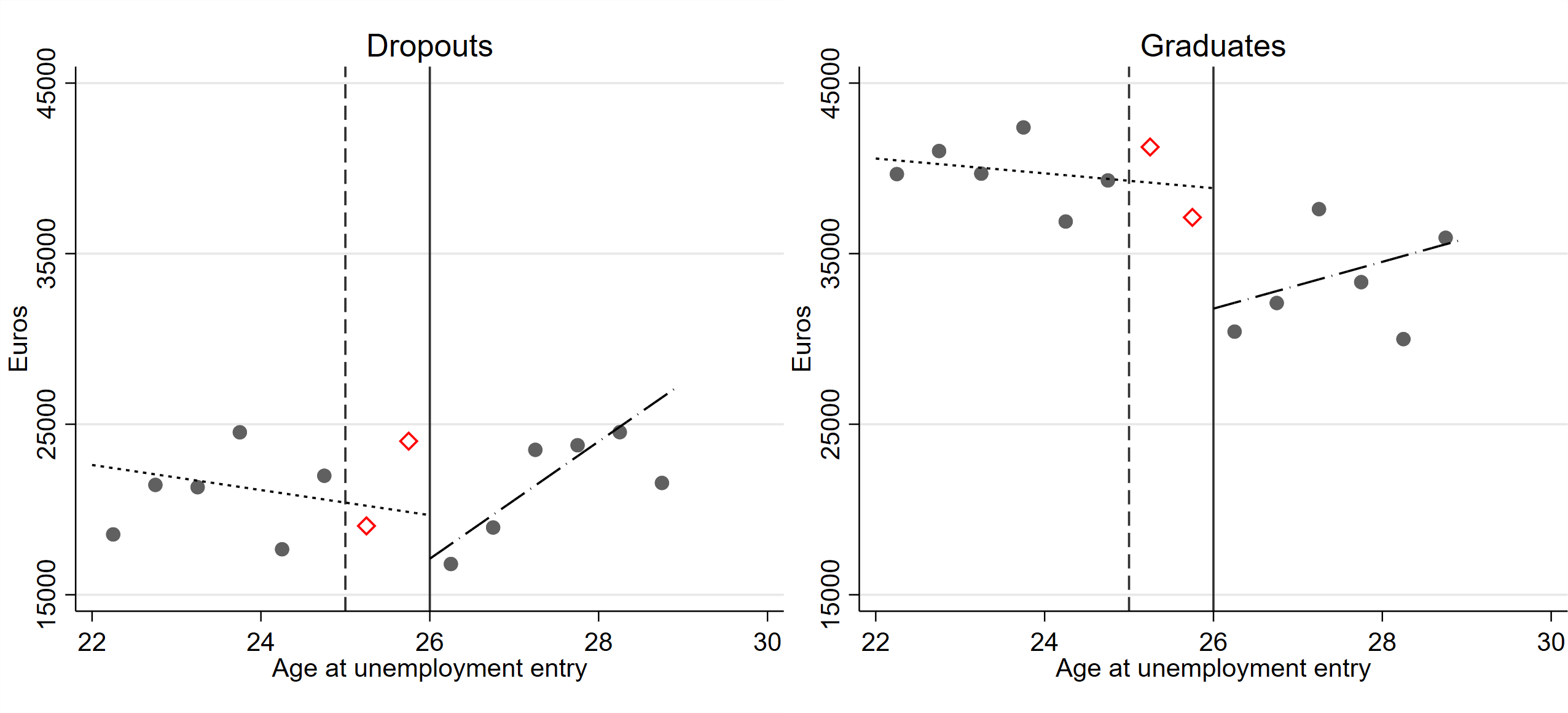}
		\caption*{\footnotesize{Note: Donut RDD estimate on the inflow sample of youths entering unemployment in 2010, using age at unemployment entry as the forcing variable with a cutoff at 26.   The outcome is the cumulative tax revenue from labor income seven years after entry into unemployment by schooling level (columns: dropouts vs. graduates), which is plotted over six age-quantile-spaced bins on each side of the donut. The two local linear splines are estimated on the reweighted observations by using the triangular kernel and the sampling weights but removing the units within the donut. We control for the set of control variables shown in Table C.2 in Online Appendix C.3.  Standard errors are clustered at the age level. The effect estimated by the donut RDD estimator at 26 years of age is 	\texteuro2,558 [--7,561; 12,677] with a p-value of 0.616 and N = 4,176 for dropouts, while for graduates it is \texteuro7,063 [306; 13,820], with a p-value 0.041 and N = 4,384.}}
	\end{figure}
	
	Even if the negative impact on non-private sector jobs offsets the positive effect on private sector employment, it is not a zero-sum game. The subsidy reinforcement increases the overall cumulative gross wage (unconditional on employment) over seven years by \texteuro5,651, although statistically insignificant (CI [--7,242; 18,544], p-value 0.385), and the paid labor income taxes, including employer and employee social security contributions, by \texteuro7,063 (CI [306; 13,820], p-value 0.041 - Figure \ref{fg:tax28un}).\footnote{We would expect the effect on gross wages to be larger than the effect on paid labor income taxes, but this inconsistency may be attributed to the large imprecision in the estimates.} This suggests that the earnings in the created private sector jobs are higher than in the substituted non-private sector jobs either because they  pay less or offer a shorter working time. Looking deeper reveals that substituted public jobs are mostly contractual jobs operated by local authorities,\footnote{See Figure A.16 in Online Appendix A.} which are, for the low-skilled population, essentially in fields such as public works, construction, electromechanics and forest work \citep{IWEPS15}. Such occupations are also in demand in the private sector, and, hence, substitutable. Unlike civil servants, who benefit from better conditions, contractual workers are employed with an open ended or fixed contract as in the private sector and often only work  part-time.\footnote{See Figure A.17 in Online Appendix A for the effect on earnings in the public sector.}

	Regarding salary conditions, there is one additional piece of evidence indicating that the additional workers in the private sector gave up low-paying jobs. For each of the public and self-employed sectors separately, we compute the RDD estimator on the average quarterly earnings over seven years (conditional on employment in one of the two sectors). These increase by \texteuro2,500 (p-value = 0.025) in public jobs,  and by \texteuro1,065 (p-value = 0.070) in self-employment.\footnote{Estimate on average earnings at 7-year distance in (i) public sector jobs: \texteuro2,500, CI [330, 4,668], p-value 0.025, N=965; and (ii) self-employment: \texteuro1,065, CI [--93, 2,224], p-value 0.071, N=552.}  This suggests that reallocated workers from unsubsidized to subsidized private firms earned less in the public (self-employment) sector than their nondisplaced counterparts. 	In the public sector these lower earnings are induced by lower hourly wages because the reduction of FTE employment in this sector is 1.9 which is very close to the reduction mentioned above of 2.1 quarters. The substituted self-employment cannot be decomposed in a similar way. We notice, however, that the gross earnings of the self-employed in the counterfactual of no subsidy reinforcement are very low: \texteuro870/quarter on average, which is in line with the evidence that in the last 20 years solo self-employment with low pay and social protection is on the rise in OECD countries as an intermediate category between employment and unemployment \citep{boeri2020solo}. Lastly, since in Subsection \ref{secpriv} we showed no effect on FTE gross wages in the private sector (conditional on private sector employment), this implies that the reallocated workers receive similar pay conditions in the private sector as their counterparts.
	
	Bringing these pieces of evidence together leads us to conclude that the reinforced hiring subsidy attract workers who would otherwise have found employment in low-paying jobs in  unsubsidized sectors. In Belgium, private sector jobs are rationed because wages are fixed in sectoral collective agreements which include wage floors that differ by sector and that are usually binding for low-skilled youths. The reinforcement of the hiring subsidy relaxes the rationing and allows low-educated youths to enter higher-paying jobs in the private sector. In the absence of the policy, most of these workers stay unemployed and would transit only after a while to a low-paying job in local public administrations and self-employment. This explains why we do not observe any substitution in the first year after unemployment entry.

	These findings suggest that private sector jobs can crowd out lower-paying local public sector jobs. This goes against the existing literature that usually finds evidence for crowding out in the other direction \citep{algan2002public, caponi2017effects, fontaine2019labour}. The fact that in our context the subsidized private sector jobs are the “good" jobs and the substituted local public sector jobs are the “bad" jobs,  while this is usually the reverse, may explain this opposite finding. The findings for high school graduates are also consistent with both signaling and human capital theory. On the one hand, the hiring subsidy reinforcement seems to provide an opportunity for some youths who would otherwise have been initially unemployed and in the long run ended up in low-paid self-employment or public sector jobs to  reveal their abilities for better paid jobs in the private sector \citep{Pallais14}. On the other hand, we cannot exclude that part of the long-run earnings gains for these workers is due to more intensive on-the-job training in these private sector positions, gradually building up thereby their earnings capacity in the long run \citep{BenPorath67, Blinder76, Mroz06}.
	
	\subsubsection{Moderation by Labor Market Tightness} \label{sec:geo}
	
	In this subsection, we investigate the moderating role of labor market tightness, which is induced by the geographic proximity to the economic hub of the Grand Duchy of Luxembourg.
	
	With a population of only 635,000 inhabitants, Luxembourg is one of the smallest countries in the European Union as well as one of the wealthiest. This is to a large extent related to the historically low corporate and personal tax rates that have attracted many multinational companies and led to the settlement of a large financial center. In this way, the country has developed into an economic hub in the region, offering more and better-paid employment opportunities. For example, in 2016 the net salary of Belgian cross-border workers aged 25-43 living in the Province of Luxembourg was 63\% higher than that of local workers \citep{Albanese23}. Due to this large economic asymmetry and the absence of language and legal barriers,\footnote{French is a common official language on both sides of the border, and the freedom of movement has existed since 1944 when the Benelux customs union was founded between Belgium, The Netherlands, and Luxembourg.} in 2010 about 38,000 workers living in Belgium crossed the border to work in Luxembourg. This represents 11\% of total employment in Luxembourg \citep{statec2}, and 25\% of employment in the nearby Belgian Province of Luxembourg \citep{Eurostat22}.

	A consequence of the proximity of such an economic attraction pole is that the labor market is much tighter close to the border with Luxembourg than farther away. To that purpose, we divide the sample based on whether the jobseekers lived either within or beyond a 60-minute driving distance from the border.\footnote{We use a 60-minute threshold since this is the observed median value in our sample. Furthermore, as shown in Figure A.18 in Online Appendix A, the share of cross-border workers decreases consistently up to 60 minutes, after which it remains flat and close to zero. Information on the average commuting time by car from the neighborhood of an individual to the closest access point in Luxembourg is retrieved from TomTom data (date of reference: 28-05-2019, arrival at 9:00 am -- \url{https://developer.tomtom.com/products/data-services}).} In 2010, the employment rate for youths aged 25-34 living closer to the border with Luxembourg was 14 percentage points higher than those living farther away (77\% versus 63\%), and the unemployment rate was only half as high (10\% versus 20\%).\footnote{The share of cross-border workers out of the total population near (far from) the border with Luxembourg was 21\% (1\%). These statistics are based on our calculations. We do not include individuals younger than 25 because a high fraction of these is still in education.}

\begin{figure}[H]
		\centering
		\caption{Discontinuity at Age 26 for the Cumulative Number of Quarters in Private Sector Employment Seven Years after Entry into Unemployment – By Distance to the Border}\label{fg:EMP28n}
		\begin{subfigure}{1\textwidth}
			\caption{Close to the border}
			\centering
			\includegraphics[width=0.84\textwidth]{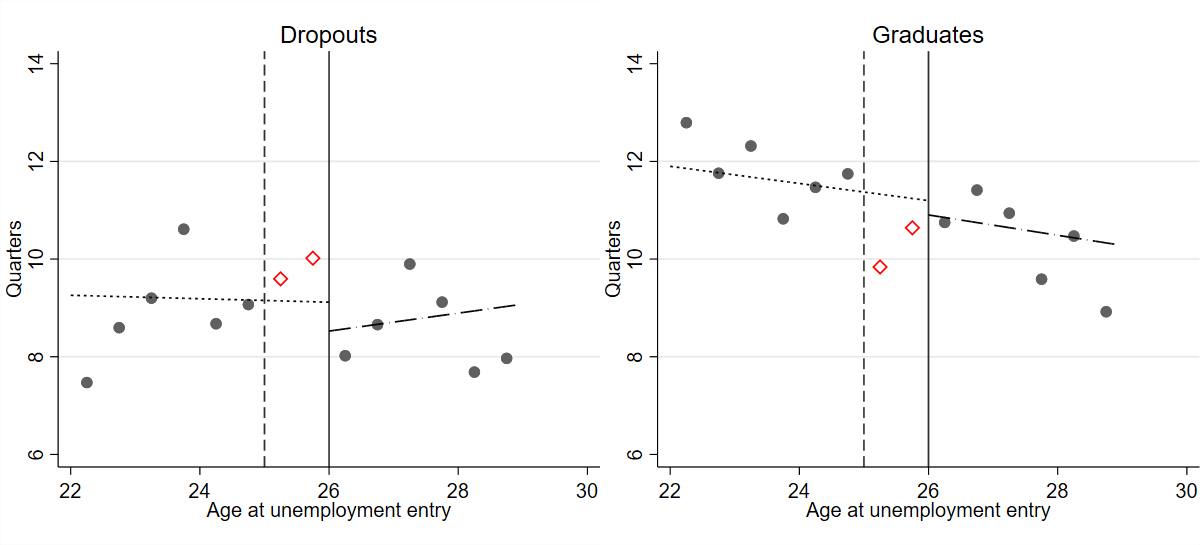}
		\end{subfigure}	
		\begin{subfigure}{1\textwidth}
			\caption{Far from the border} 
			\centering
			\includegraphics[width=0.84\textwidth]{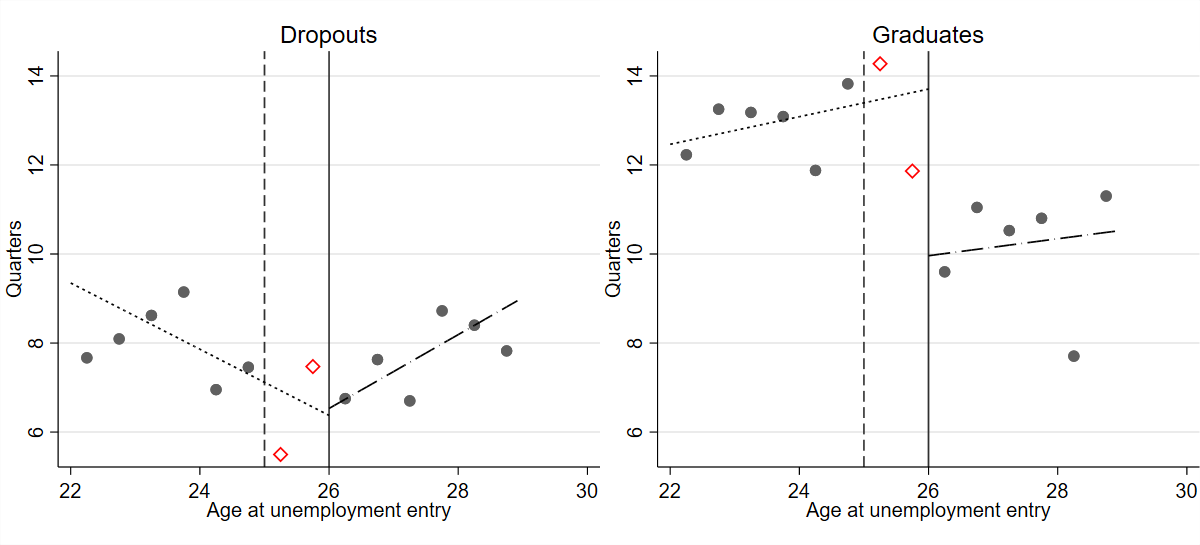}
		\end{subfigure}	
		\caption*{\footnotesize{Note: Donut RDD estimate on the inflow sample of youths entering unemployment in 2010, using age at unemployment entry as the forcing variable with a cutoff at 26.   The outcome is the cumulative number of quarters employed in private sector employment seven years after entry into unemployment by schooling level (dropouts vs. graduates) by driving distance to the border: below one hour (a) or above one hour (b), which is plotted over six age-quantile-spaced bins on each side of the donut. The two local linear splines are estimated on the reweighted observations by using the triangular kernel and the sampling weights but removing the units within the donut. We control for the set of control variables shown in Table C.2 in Online Appendix C.3.  Standard errors are clustered at the age level. The effect estimated by the donut RDD estimator at 26 years of age is for dropouts (graduates)  close to the border: 0.6 quarters [--2.0; 3.2] with		a p-value of 0.653 and N = 1,443 (+0.3 quarters [--1.9; 2.4], p-value 0.786 and N = 1,939), while for dropouts (graduates) far from the border it is --0.2 quarters [--3.3; 3.0] with	a p-value of 0.921 and N = 2,636 (3.7 quarters [0.7; 6.8], p-value 0.016 and N = 2,432).}}
	\end{figure}

	We then look at the differential amount of subsidy received for hiring jobseekers at different border distance. We estimate that close to the border high school dropouts (graduates) finding a job within one year after unemployment entry are entitled  additionally to \texteuro414 (\texteuro300) due to the subsidy reinforcement.\footnote{See Figure A.19 in Online Appendix A.} Afterwards, we display the long-run evolution of the RDD effect on the number of quarters in private sector employment in Figure \ref{fg:EMP28n} panel (a). It can be seen that this effect is not significantly different from zero. This means that close to the border the hiring subsidy is, as expected, a complete deadweight. In contrast, for jobseekers living further away, the subsidy reinforcement grants \texteuro249 (\texteuro377) more to the dropouts (graduates). This raised, for graduates only, employment by 3.7 quarters seven years after entry into unemployment (panel (b)). This represents a proportional increase of 38\% relative to those  older than 26, and larger than the overall effect for the full population reported in Figure \ref{fg:EMP28}.\footnote{In Figures A.20-A.21 in Online Appendix A, we present the evolution of the RDD effects over seven years after entry into unemployment by border distance.}
	
	We also report the effect on the accumulated time spent in private sector employment estimated from a model in which we interact the splines and the treatment indicator of the donut RDD estimator with the travel distance from the border with Luxembourg, instead of splitting the sample into subgroups.\footnote{See Figures A.22 and A.23 in Online Appendix A.} By predicting  the effect over the distance for high school graduates, in linear and quadratic specifications, we see that the treatment effect becomes significant only from about 40 minutes from the border. Below this threshold, the effect is never significantly different from zero. Above it continues to increase, but the quadratic specification shows that it levels off beyond 60 minutes from the border. The corresponding effects for high school dropouts confirms that for this group, the effects are close to zero for any travel distance.\footnote{We also estimate a similar interactive model for high school graduates with the number of quarters of cross-border work 7 years after unemployment as the outcome. As shown in Figure A.24 in Online Appendix A, no significant reduction in cross-border work is found.}
	
	The Belgian hiring subsidy does not induce firms to create new jobs close to the border with Luxembourg because most of the productive workforce is already employed or prefers working in that country. On the Belgian side of the border, vacancies are to a large extent for replacement hiring in essential occupations and not for new job creation. In our data the share of Belgian private sector employment out of the total employment is smaller close to the border with Luxembourg than farther away\footnote{This was 42\% (61\%) within (beyond) 60 minutes driving distance from the border for workers aged 25-34.} and those jobs tend to be in low-status occupations.\footnote{Our data show that near (far from) the border, 47\% (38\%) of jobs in the private sector are blue-collar jobs, 32\% (26\%) are part-time jobs, the average gross full-time daily salary is \texteuro100 (\texteuro104), 42\% (32\%) of jobs are in firms with fewer than 20 employees, and 16\% (28\%) are in firms with more than 500 employees.} These stylized facts suggest that local employment is indeed in essential occupations, allowing those at home to get what they need day-to-day and for which labor demand is relatively inelastic.

	Kline and Moretti (2013) argue that in a tight labor market where there is excessive job creation, subsidizing hires is inefficient because vacancies crowd each other out. We are the first to demonstrate empirically that labor market tightness can moderate the effectiveness of hiring subsidies, even if vacancy creation is across the border in firms that are not eligible for the hiring subsidy. The reason is that in the absence of mobility barriers, the labor market tightness in Luxembourg extends across the border into Belgium. 
	
	\subsection{Additional Analyses}\label{sec:robustness}
	
	\subsubsection{Sensitivity tests}\label{sec:sens}
	In this section, we report validation tests.\footnote{See Figures A.25-A.41  in Online Appendix A and Table H.1 in Online Appendix H.} First, we rerun the one-sided donut RDD estimator widening or narrowing the bandwidth. The results are close to the benchmark estimates. Second, we remove the conditioning variables from the RDD estimator and again obtain similar estimates. Third, we test the sensitivity of the results on the effect of the subsidy near the border by reducing the distance to 45 or 30 minutes by car, which show very similar results. Finally, we let the spline on the right of the donut predict the outcome inside the “hole” and estimate the treatment effect at age 25. Estimates are similar to those for individuals aged 26.
	
	We also implement three placebo tests for the donut RDD estimator. First, we estimate whether we detect any statistically significant jump at age 26 for individuals entering unemployment before the introduction and after the abolition of the Win-Win plan (2008 and 2012). Second, we check whether at the age of 26 we find a significant discontinuity in the outcomes for the unemployed with a tertiary degree, who were not eligible for the Win-Win subsidy. Third, we implement several placebo tests that use false cutoff points of the forcing variable. Fourth, we apply the donut RDD estimator to detect jumps in the control variables at the discontinuity. These placebo tests deliver insignificant estimates. Finally, we implement the density test done in the empirical literature \citep{Cattaneo20} with or without the donut. Both tests show a non-significant change of density at the discontinuity. Overall, all of these validation tests confirm the reliability of the treatment effect we have found for our treated population.
	
	\subsubsection{Difference-in-differences}\label{sec:estimatorDD}			
	
	In this subsection, we exploit the time variation in eligibility introduced by the implementation of the Win-Win reinforcement plan in 2010 and its subsequent abolition in 2012. This approach allows us to control for time-invariant unobserved heterogeneity between youths below and above age 26, under the parallel trend assumption of the difference-in-differences (DiD) estimator. This validation analysis confirms that the subsidy reinforcement has a long-term impact exclusively for high school graduates who entered unemployment during the reform period; this impact is not observed in other entry cohorts, either before or after the reform.\footnote{This specificity is reassuring considering that business cycle conditions in 2008 and 2012 were significantly different. If our findings were merely artifacts of these varying business cycle conditions, then the long-term effects for these two placebo cohorts (pre- and post-reform) would also be expected to differ.}  Moreover, the effect does not extend to other educational groups, such as high school dropouts and ineligible post-secondary graduates, which we use as a placebo group to test for the presence of age-specific time trends. Additionally, the magnitude and evolution of these long-term effects for high school graduates closely align with those identified in the RDD, further enhancing the credibility of our findings as discussed in the main results section.

\begin{figure}[H]
		\centering
		\caption{Evolution over Entry Cohorts and Schooling Level of the Cumulative Number of Quarters in Private Sector Employment} \label{afg:DDtrendEMP}	
		\begin{subfigure}{0.428\textwidth}
			\caption{Dropouts} 
			\centering\includegraphics[width=1\textwidth]{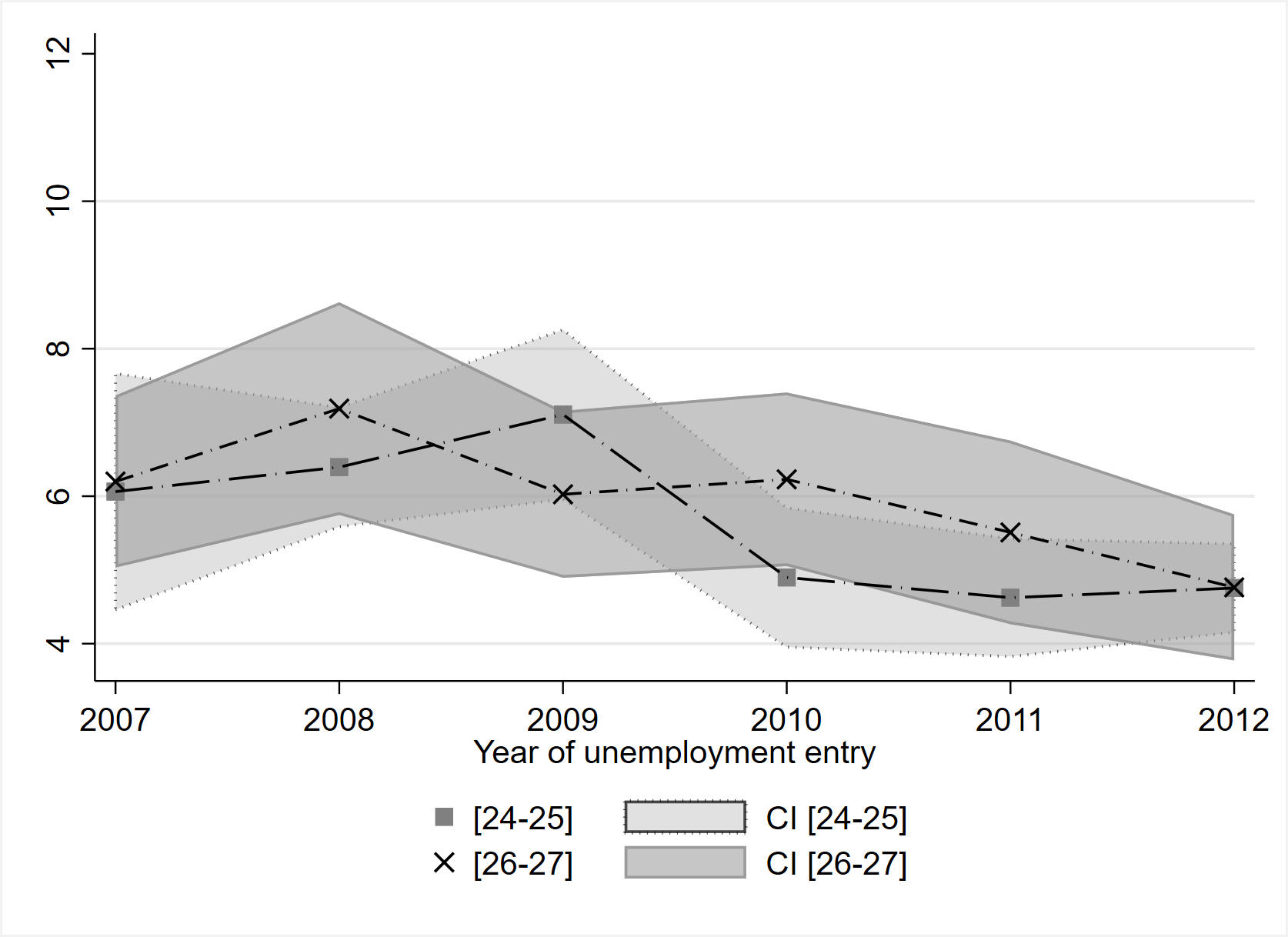}
		\end{subfigure}	
		\begin{subfigure}{0.428\textwidth}
			\caption{Graduates} 
			\centering\includegraphics[width=1\textwidth]{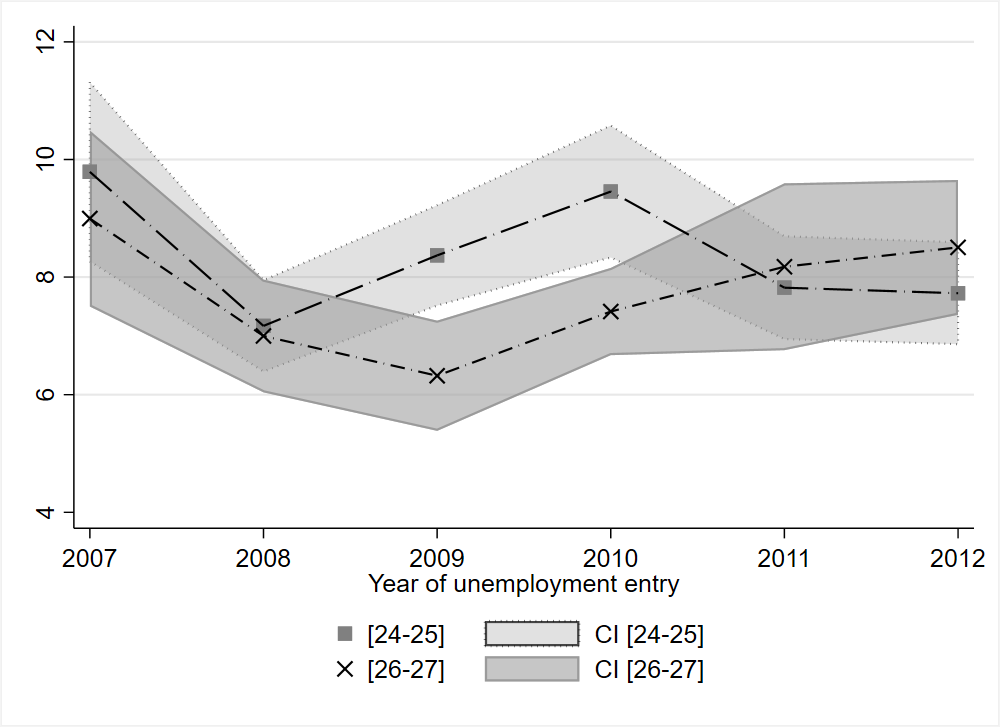}
		\end{subfigure}	
		\caption*{\footnotesize{Note: Evolution of the cumulative outcomes measured 5 years after entry into unemployment  on the cumulative number of quarters in private sector employment  by entry cohort and schooling level: dropouts (left) vs. graduates (right). The treated are aged 24-25 at unemployment entry, while controls are aged 26-27. The 2009 cohort is not used as control period since jobseekers registering in 2009 quickly enter the subsidized treatment period. Data are reweighted by the sampling weights.}}	
	\end{figure}

	First, Figure \ref{afg:DDtrendEMP} shows the evolution of cumulative number of quarters in private sector employment five years after entry into unemployment by entry cohort.\footnote{We stop the outcome at 5 years after entry into unemployment because this is the maximum observational period for the 2012 entries. See Figures A.42 and A.43 in Online Appendix A for the other cumulative outcomes.} We do not observe any statistically significant difference between jobseekers aged 24-25 (treated) and 26-27 (controls) entering unemployment during either the pre-treatment period (2007 and 2008) or the post-intervention period (2012). For the entry cohorts of 2009 (treated in 2010) and 2010, we observe a jump in the outcomes only for the treated high school graduates, which reabsorbs when the subsidy is  phased out in 2011. 
	
	Second, we look at the evolution over three entry cohorts (2008, 2010 and 2012) and three education groups (high school dropouts, high school graduates, and post-secondary graduates, which we use as a placebo group) of the cumulative number of quarters in private sector employment, varying the elapsed time since entry into unemployment. Among all entry cohorts and education groups, we observe a significant difference only for the 2010-cohort of high school graduates. This finding remains robust to an IPW estimator controlling for compositional differences in observable characteristics.\footnote{See Figures A.44-A.45 in Online Appendix A.} 
	
\begin{figure}[H]
		\centering
		\caption{Evolution of the DiD Effect on Cumulative Number of Quarters in Private Sector Employment Over Calendar Time at Registration} \label{afg:dd}	
		\begin{subfigure}{0.6\textwidth}
			\centering\includegraphics[width=1\textwidth]{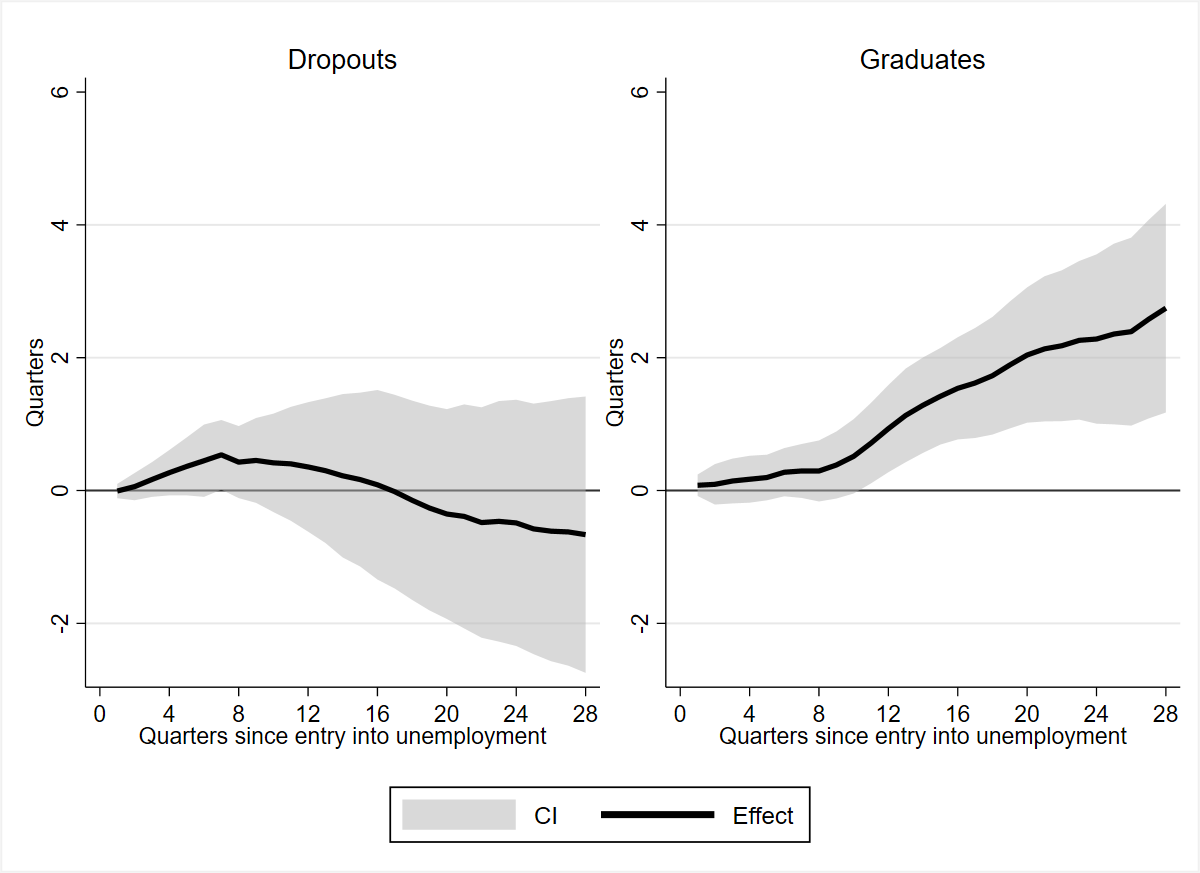}
		\end{subfigure}	
		\caption*{\footnotesize{Note: Evolution of the effect estimated with a doubly robust DiD estimator \citep{SANTANNA20} and confidence intervals (CI) for  the cumulative number of quarters in private sector employment since unemployment and by schooling level: dropouts (left) vs. graduates (right). The DiD estimator is implemented for each year after entry into unemployment until 7 years later. The treated are aged 24-25 at unemployment entry, while controls are aged 26-27. Units registering in (2008) 2010 are considered in the (pre-) treatment period. Data are reweighted by the sampling weights. We control for the set of control variables shown in Table C.2 in Online Appendix C.3.  Standard errors are clustered at the age level. For dropouts (graduates), the effect at 7 years is -0.7 quarters [--2.7; 1.4] with a p-value of 0.532 and N = 1,942 (+2.7 quarters [1.2; 4.3], p-value 0.001 and N = 1,839).}}	
	\end{figure}

	Finally, we formally estimate the treatment effect by implementing the doubly robust DiD estimator as explained in Section \ref{sec:DiD1} and Online Appendix E. The results closely resemble those obtained by implementing the RDD estimator and are displayed in Figure \ref{afg:dd}. Importantly, the DiD estimator also replicates the effects on other outcomes and the heterogeneous response found for individuals living near or far from the border with Luxembourg.\footnote{See Figures A.46-A.48 in Online Appendix A.} Validation analyses on the trends during the pre-treatment periods and the unaffected placebo group of post-secondary graduates are presented in Online Appendix E.1.
	
	\subsubsection{Cost-Benefit Analysis}\label{sec:cba}			
	
	In this final section, we implement a cost-benefit analysis of the subsidy reinforcement following the marginal value of public funds (MVPF) framework proposed by \citet{Hendren20} and designed for measuring the long-run effectiveness of policies. The MVPF is the ratio of the beneficiaries’ marginal willingness to pay (WTP) for the reinforcement of the
	hiring subsidy to the net marginal cost (NC) to the government of this policy inclusive of any behavioral impact on the government budget. The MVPF can range from negative to positive values. It is convention to set it to plus infinity ($+\infty$) whenever the net cost to the government is negative and the WTP positive because in this case the policy has value (WTP$>0$) and finances itself (NC$<0$). We refer to Online Appendix G for details.
	
	Overall, the Win-Win plan did not seem to impose much cost on the government, and for high school graduates, the reform may even have paid for itself, which is driven by the higher tax contributions. The overall MVPF for dropouts is 4, while for the graduates, it is $+\infty$. The policy is also more likely to be self-financing in areas of low labor market tightness. The point estimates of the MVPF are equal to $+\infty$ and $1$, respectively. The finding that the hiring subsidy could be self-financing aligns with \citet{Cahuc19}, who report that the short-run net cost per created job of the hiring credit during the Great Recession in France equals zero. Our evidence indicates that, in the long run, the net cost to the government even becomes negative. A word of caution on these results, however, is needed as the confidence intervals of the MVPF are very wide. They always encompass zero and plus infinity, implying that we require a larger sample to make this cost-benefit analysis informative for any policy conclusions.
	
	\section{Conclusion}\label{sec:conclusions}
	
	In this paper, we evaluate the employment effects of a temporary reinforcement of a hiring subsidy targeted at low- and medium-skilled unemployed youths during the recovery from the Great Recession in Belgium. A primary objective of this paper is to uncover to what extent such targeted and temporary hiring subsidies can be effective in reversing the long-term scarring effects that recessions can have on young workers. We contribute to the existing literature by focusing on long-term effects, considering the substitution of low-paying local public jobs and self-employment with subsidized private sector jobs, and studying the moderating effects of labor market tightness. The mentioned moderation could be identified by the specific sample that was drawn from a region close to the border with Luxembourg, a prosperous economic hub that attracts substantial cross-border work from Belgium. The main causal analysis exploits an eligibility age cutoff of 26 years for the hiring subsidy and is based on a one-sided donut regression discontinuity design \citep{Gerard21} to estimate the intention-to-treat effect. The qualitative findings are robust to using an alternative identification strategy, i.e., the doubly robust semi-parametric difference-in-differences method of \citet{SANTANNA20} with treatment and control groups defined closely around the aforementioned age cutoff.
	
	We show that the subsidy reinforcement accelerates job-finding in the short run by about 10 percentage points for both skill-level groups. The subsidy generates persistent employment effects exclusively for high school graduates. Seven years after entry into unemployment, high school  graduates have accumulated about three quarters more employment in the private sector than in the counterfactual of eligibility for a substantially lower hiring subsidy. While employment gains in the short run stem from a reduction in unemployment, in the longer term, these are primarily due to a decrease in low-paying employment in the unsubsidized sectors, such as self-employment and public sector jobs. However, this substitution creates better career prospects for high school graduates who might otherwise be trapped in low-paid jobs. Our analysis also reveals that the tight labor market induced by the presence of the neighboring employment hub of Luxembourg across the border results in a complete deadweight loss for the creation of private sector jobs in an area near the border. The cost-best analysis suggests that the reinforcement of the hiring subsidy could pay for itself, especially for high school graduates living far from the border with Luxembourg. But this finding requires corroboration because of the lack of precision of the estimates emerging from this analysis.
	
	Our results imply that targeting a pure hiring subsidy at high school dropouts during the recovery from a recession can at most accelerate the transition to temporary jobs and cannot persistently improve the labor market position of this group. The absence of such effects might be linked to the short duration and the low skill requirements of jobs that are available for dropouts. A minimum skill level seems to be a condition for the effectiveness of “work first” policies. For high school graduates, our policy conclusions are more positive. Even if the reform did not increase overall employment, it did open up the opportunity for youths to substitute in the long run better-paid jobs in the private sector for low-paid jobs in the local public sector and self-employment. Nevertheless, policymakers should be aware that these beneficial effects are only evident in slack labor markets. Furthermore, they should acknowledge that the inefficacy associated with labor market tightness, in the absence of mobility restrictions, can be influenced by conditions across the border. In a scenario marked by pressure on public spending, identifying factors that optimize the cost-effectiveness of public policies should be a top priority for both policymakers and researchers.

	\clearpage
	
	\addcontentsline{toc}{section}{References}
	\setlength{\bibsep}{1.5pt}
	\bibliographystyle{apalike}
	
	\bibliography{references10}

\end{document}

%% file: descriptives1.tex
{
\def\sym#1{\ifmmode^{#1}\else\(^{#1}\)\fi}
\begin{tabular}{l*{4}{c}}
\hline
                              &\multicolumn{2}{c}{High school}&\multicolumn{2}{c}{High school}\\\
                              &\multicolumn{2}{c}{dropouts (22-25)}&\multicolumn{2}{c}{graduates  (22-25)}\\\cmidrule(lr){2-3}\cmidrule(lr){4-5}
                              &\multicolumn{1}{c}{\shortstack{All\\(1)}}&\multicolumn{1}{c}{\shortstack{Win-Win\\(2)}}&\multicolumn{1}{c}{\shortstack{All\\(3)}}&\multicolumn{1}{c}{\shortstack{Win-Win\\(4)}}\\
\hline
Take-up Win-Win during any month within 1 year      &        0.19&        1.00&        0.20&        1.00\\
                              &      (0.39)&      (0.00)&      (0.40)&      (0.00)\\
Employed in the private sector at the end of any quarter within 1 year     &        0.44&        0.92&        0.58&        0.93\\
 &      (0.50)&      (0.27)&      (0.49)&      (0.25)\\
Total quarters in the salaried private sector in 7 years &        8.17&       12.24&       12.54&       16.74\\
    &      (8.87)&      (8.58)&      (9.86)&      (8.82)\\
Total quarters in other employment in 7 years        &        2.47&        2.24&        4.74&        2.50\\
       &      (5.54)&      (4.55)&      (7.95)&      (5.86)\\
\hline
N                             &        2209&         394&        2838&         520\\
\hline
\end{tabular}
}